\documentclass[article,prc,flaotfix,nofootinbib,aps]{revtex4}
\usepackage{amsmath}
\usepackage{amsfonts}
\usepackage{amssymb}
\usepackage[english]{babel}
\usepackage{graphicx}
\usepackage{color} 
\usepackage{dcolumn} 

\newcommand{\jpsi}{$J/\psi$}
\newcommand{\SigAbs}{$\sigma_\textrm{abs}$}
\newcommand{\Sabs}{$S_\textrm{abs}$}
\newcommand{\Sshad}{$S_\textrm{shad}$}

\newcommand{\pt}{$p_T$}
\newcommand{\sNN}{$\sqrt{s_{_{NN}}}$ }
\newcommand{\pp}{$pp$}
\newcommand{\dNdch}{$dN_\textrm{ch}/dy$}
\newcommand{\SdNdch}{$1/S_\perp dN_{\textrm{ch}}/dy$}
\newcommand{\Rcp}{$R_{\rm CP}$}
\newcommand{\Nc}{$N_c$}
\newcommand{\bb}{$b$}

\newcommand{\SpsiTtSs}{$(S^{\textrm{TT}}/S^{\textrm{SS}})_{J/\psi}$}
\newcommand{\dEdy}{$dE_T/dy$ }

\newcommand{\RUAu}{$R^{\textrm{UU}}_{\textrm{AuAu}}$}
\newcommand{\Ups}{$\Upsilon$}

\begin{document}

\title{Prospects for quarkonia production studies in U+U collisions}

\author{Daniel Kiko\l{}a$^1$\footnote{Present address: Physics Department, Purdue 
University, West Lafayette, IN 47907}, Grazyna Odyniec$^1$, Ramona Vogt$^{2,3}$}
\affiliation{$^1$Nuclear Science Division, Lawrence Berkeley National 
Laboratory, Berkeley, CA 94720 \\
$^2$Physics Division, Lawrence Livermore National Laboratory, Livermore,
CA 94551 \\
$^3$Physics Department,
University of California, Davis, California, 95616}
\date{\today}

\begin{abstract}
Collisions of deformed uranium nuclei provide a unique opportunity to study 
the spatial dependence of charmonium in-medium effects. By selecting the 
orientations of the colliding nuclei, different path lengths through the 
nuclear medium could be selected within the same experimental environment. 
In addition, higher energy densities can be achieved in U+U collisions relative
to Au+Au collisions.  In this paper, we investigate the prospects for charmonium
studies with U+U collisions.  We discuss the effects of shadowing and nuclear 
absorption on the \jpsi\ yield.  We introduce a new observable which could 
help distinguish between different types of \jpsi\ interactions in hot and 
dense matter.   
\end{abstract}

\maketitle

\section{Introduction}
\label{Sec:Intro}

Collisions of deformed nuclei, such as $^{238}$U, are an interesting alternative 
to those of more spherical $^{197}$Au nuclei at RHIC because changes in the
nuclear orientation allow wider variations in energy density within the same 
system \cite{Heinz:2004ir,PhysRevC.73.034911,Masui2009440,Hirano:2010jg}. 
Previously, studies of U+U 
collisions were discussed in the context of elliptic flow. The quantity 
$v_2/\epsilon$, where $v_2$ is the elliptic flow parameter and $\epsilon$ is 
the initial spatial eccentricity, provides valuable information about the 
matter created in heavy-ion collisions~\cite{Voloshin:1999gs}. In a dilute 
system, $v_2/\epsilon$ scales like \SdNdch, where \dNdch\ is the charged 
particle density and $S_\perp$ is the transverse area of the overlap zone. 
In central Au+Au collisions at \sNN = 200 GeV $v_2/\epsilon$ reaches 
the limit predicted by ideal hydrodynamics for strongly-interacting matter
\cite{STAR:Flow} which can be interpreted as a signature of quark-gluon plasma 
(QGP) formation. It was argued that U+U interactions could provide higher 
densities and a larger spatial eccentricity, testing whether 
$v_2/\epsilon$ saturates, as predicted by ideal hydrodynamics, or increases 
further with \dNdch\ \cite{Kolb:2000sd,Heinz:2004ir,Hirano:2010jg}.

Charmonium production is another key observable for studying the properties of 
the hot and dense matter created in relativistic heavy-ion collisions. 
Some time ago \jpsi\ suppression was proposed as a signal of QGP 
formation~\cite{Matsui1986416}.  The suppression is expected to arise from 
color screening of the binding potential in a QGP, similar to Debye screening 
in a classical electromagnetic plasma. The magnitude of the suppression 
depends on the charmonium binding energy and the energy density of the medium, 
related to the temperature. Therefore, studies of \jpsi\ production, 
particularly at low transverse momenta, reveal the thermodynamic properties 
of the medium.  

The suppression of \jpsi\ production has been studied in detail at the CERN
SPS and at RHIC. The NA50 and NA60 experiments at the CERN SPS, \sNN = 17.3 GeV,
observed strong \jpsi\ suppression as a function of collision 
centrality~\cite{NA501997,NA50PbPb,Na602007}. Results from the 
PHENIX collaboration at RHIC show that 
the \jpsi\ suppression in midrapidity Au+Au collisions at \sNN = 200 GeV
is similar to that observed at SPS energies~\cite{PhenixJpsiAuAu} even though 
the energy densities and temperatures reached at RHIC are much higher than 
those at the SPS. Moreover, the forward rapidity \jpsi\ suppression at RHIC is 
stronger than at midrapidity. Such a pattern suggests that either additional 
processes compensate for the effects of color screening or the suppression has 
different origins at the two energies.  

There are several processes which 
may affect charmonium production in nucleus-nucleus collisions.  
For comprehensive reviews of different 
aspects of quarkonium in-medium interactions, see 
Refs.~\cite{Kluberg:2009wc,Rapp:2008tf}. Here we briefly summarize a few 
relevant features, some of which we address in more detail later.

Suppression by cold nuclear matter (CNM) effects such as parton shadowing and 
absorption by nuclear matter are very important, although not yet established 
with satisfactory precision \cite{RVgeneral}. For example, parameterizations 
of gluon shadowing differ significantly between models \cite{EPS09}. At RHIC 
energies, CNM effects are studied 
in d+Au collisions. Changing the shadowing parametrization can significantly
affect the magnitude of the absorption cross section required to match the
measured nuclear modification factor $R_{\rm dAu}$. In addition, feed-down 
from radiative or hadronic decays
account for 40-50\% of the observed $1S$ quarkonium states 
(\jpsi\ and $\Upsilon$). The higher mass quarkonium states have very 
different radii and formation times and should thus also have different 
absorption cross sections.  While the experimentally measured $\psi'$ absorption
cross section is significantly larger than the $J/\psi$ at fixed-target 
energies, the difference decreases with increasing \sNN \cite{NA50,E866}.  The
energy and $A$ dependence of $\chi_c$ absorption is unknown \cite{RVabschi}.
The importance of absorption for higher bottomonium states is also not known.

If color screening dominates the \jpsi\ in-medium interaction, the 
characteristic pattern of sequential charmonium ``melting" should be observed. 
The binding energies of higher mass charmonium states, $\psi'$ and $\chi_c$, 
are smaller than the \jpsi.  Therefore these 
states dissociate at lower energy densities compared to \jpsi\ so that, at 
higher $T$ their feed-down would no longer contribute to the measured \jpsi\ 
yield. The remaining, direct, \jpsi\ 
production would be suppressed at larger energy densities. This process would 
result in a step-like dependence of the \jpsi\ survival probability as a 
function of energy density \cite{Karsch:2005nk}.
A similar sequential melting would apply to bottomonium.

Suppression due to interactions with comoving partonic and hadronic matter 
has also been postulated. The $A$ dependence of comover 
dissociation is similar to that
of nuclear absorption \cite{RVSG,RVPhysRept}.  The effects are generally 
assumed to be small although they strongly depend on the model 
\cite{RVSG,Capella,HSD}.
 
\jpsi\ production may be also enhanced by statistical 
coalescence in QGP. In this case, the \jpsi\ yield, $N^{J/\psi}_{\textrm{stat}}$, is 
proportional to the square of the number of charm quarks in the system, \Nc, so
that $N^{J/\psi}_{\textrm{stat}} \propto N^2_\textrm{c}$. 
The charm yield is expected to be proportional to the number of binary 
$NN$ collisions, $N_{\rm bin}$, i.e. $N_{c \overline c} \propto N_{\rm bin}$.  The
$c \overline c$ production rate is not expected to be modified in the QGP. 
If the density of charm quarks is high enough, then secondary \jpsi\ production
by coalescence of uncorrelated $c$ and $\overline c$ quarks at hadronization
can occur \cite{ThewsMangano,statistical}.
 
Combinations of the above effects have previously been used to describe 
the \jpsi\ data. Despite different assumptions about the initial state of the
system, models including cold matter effects together with different 
combinations
of color screening, comover absorption, and coalescence give quantitatively similar 
results at RHIC \cite{RVTFTU}. It is expected that U+U collisions could provide 
additional means for distinguishing between these effects.

By selecting particular orientations of U+U collisions, different path lengths 
through nuclear matter, resulting in different effective nuclear 
absorption cross sections and available energy densities, could be studied. 
Interactions with the long axes of U nuclei 
aligned along the beam (``Tip+Tip" or ``TT" configurations) give the highest 
energy densities as well as the longest path, $L$, through the matter. In 
contrast, $L$ is shortest in the configuration where the short axes of the U 
nuclei are aligned along the beam axis (``Side+Side" or ``SS" configurations). 
Different orientations of U+U interactions could study nuclear absorption with 
reduced uncertainties due to shadowing, as we will show. In general, 
U+U interactions provide an additional check on models which 
describe the Au+Au data because some effects, such as shadowing, 
should be similar in U+U and Au+Au interactions.

We investigate the feasibility of complementing Au+Au collisions with U+U
studies of charmonium production and suppression to distinguish between 
scenarios of \jpsi\ in-medium interactions.  We also discuss CNM effects in 
U+U collisions. 

\section{Computational framework}

Here, we discuss two types of cold nuclear matter effects: nuclear absorption 
and shadowing.  Shadowing, the modification of the parton densities in nuclei 
relative to free protons, is an initial-state effect.  Nuclear absorption,
the breakup of the charmonium state as it traverses nuclear matter, is a 
final-state process. The two effects are assumed to factorize.
We neglect interactions with hadronic comovers. 

We employ the computational framework for studies of 
cold nuclear matter effects 
described in Ref.~\cite{Vogt:2010aa}.  The nuclear parton densities, 
$f_i^A(x,Q^2,\vec{r}_{T},z)$, where $A$ is the atomic mass number, $\vec{r}_{T}$ and $z$ are the transverse and longitudinal location of the
parton in position space, $x$ is the parton momentum fraction and 
$Q^2$ is the interaction scale, factorize into
the nucleon density in the nucleus, $\rho_A(\vec{r}_{T},z)$, independent of the 
kinematics; the nucleon parton density, $f_i^p(x,Q^2)$, independent of $A$;
and a shadowing ratio, $S^i(A,x,Q^2,\vec{r}_{T},z)$ that
parameterizes the modifications of the nucleon parton densities in the nucleus
so that
\begin{eqnarray}
f_i^A(x,Q^2,\vec{r}_{T},z) & = & \rho_A(\vec{r}_{T},z) 
S^i(A,x,Q^2,\vec{r}_{T},z) f_i^p(x,Q^2) \, \, .
\label{fanuc} 
\end{eqnarray}
The effect of shadowing is stronger in central collisions, while at asymptotic
distances shadowing disappears.  Averaging over impact parameter, $b$, the minimum
bias result measured in nuclear deep-inelastic scattering is regained.
We use the EPS09 shadowing parametrization \cite{EPS09} and assume that the
impact parameter dependence of shadowing is proportional to the local 
nuclear density \cite{SpatialShadDep1999} and exhibits a relatively sharp 
transition region around $b \sim 2R_A$, see 
Fig.~\ref{Fig:Jpsi_surv_probab_shadowing}.   If one assumes instead that the
impact parameter dependence is proportional to the path length $L$, $S^i$
exhibits a weaker transition region \cite{Inhomogeneous:Shadowing}. 
We discuss the consequences of different assumptions of this $b$ dependence
in the next section.  At RHIC 
energies quarkonium production is dominated by gluon fusion.  Therefore, 
shadowing generally refers to gluon shadowing in this context.
We employed the Woods-Saxon density distribution for deformed nuclei 
\cite{Masui2009440} to describe $\rho_{\rm U}(\vec{r}_{T},z)$,
\begin{eqnarray}
\rho_{\rm U}(\vec{r}_{T},z) = \rho_0 (1 
+ \exp[(r - R_C)/a])^{-1} \\
R_C \equiv R_0(1 + \beta_2 Y_{20} + \beta_4 Y_{40})
\end{eqnarray}
where $\rho_0$ is the density at the center of the nucleus,
$R_C$ is the total effective radius, $Y_{20}$ and $Y_{40}$ are 
Legendre polynomials, dependent on 
$\cos\theta$, where $\theta$ is the polar angle, $\vec{r}_{T} = (r_{T} \cos \theta, r_{T} \sin \theta)$ and $r = \sqrt{|\vec{r}_{T}|^2 + z^2}$.  The deformation 
parameters are $\beta_2 = 0.28$ and $\beta_4 = 0.093$ with 
$R_0 = 6.81$~fm and $a = 0.54$~fm.  In the SS orientation, $\cos\theta = 0$ and
$R_C = R_{\rm S} = 6.41$~fm while in the TT orientation, $\cos\theta = 1$ and 
$R_C = R_{\rm T} = 8.56$~fm.

To implement nuclear absorption on quarkonium
production, the production cross section 
is weighted by the
survival probability, $S_{\rm abs}$, 
\begin{eqnarray} 
S_{\rm abs}(\vec b - \vec{r}_{T},z^\prime) = \exp \left\{
-\int_{z^\prime}^{\infty} dz^{\prime \prime} 
\rho_A (\vec b - \vec{r}_{T},z^{\prime \prime}) 
\sigma_{\rm abs}(z^{\prime \prime} - z^\prime)\right\} \, \, 
\label{nsurv} 
\end{eqnarray}
where $z^\prime$ is the longitudinal production point and 
$z^{\prime \prime}$ is the point at which the state is absorbed. 
The nucleon absorption cross section, $\sigma_{\rm abs}$, typically 
depends on the spatial location at which the
state is produced and how far it travels through the medium.

\section{Cold nuclear matter effects in U+U collisions}
\label{Sec:CNM}

In this section, we investigate whether it is possible to separate the CNM 
effects of shadowing and absorption in U+U collisions.  Nuclear absorption 
depends on the nuclear path length $L$.  Therefore, it is expected to be 
considerably different in the TT and SS orientations of U+U collisions.  Because
$L$ also depends on the impact parameter, \bb, the effective magnitude of 
nuclear absorption also changes with \bb. The dependence of shadowing on the TT and SS orientations 
is expected be less significant. 
Moreover, if the difference between collisions in the TT and SS orientations 
does not depend (or depends only weakly) on the strength of the shadowing, 
then the ratio of quarkonium production in TT and SS orientations  would be 
effectively independent of the shadowing parametrization. This would greatly 
simplify the extraction of absorption effects because the systematic error 
due to shadowing would be significantly reduced.

We first investigate shadowing and nuclear absorption separately and then 
discuss the combined intensity of these CNM effects in U+U collisions. 
In order to estimate 
the strength of each effect, we compare results
in the TT and SS orientations where the difference between the two cases 
should be most apparent.  We assume that it will be possible to select these
configurations with an efficiency appropriate for charmonium studies, as
discussed later.

\begin{figure}[htdp]
\begin{tabular}{cc}
\includegraphics[width=0.48\textwidth]{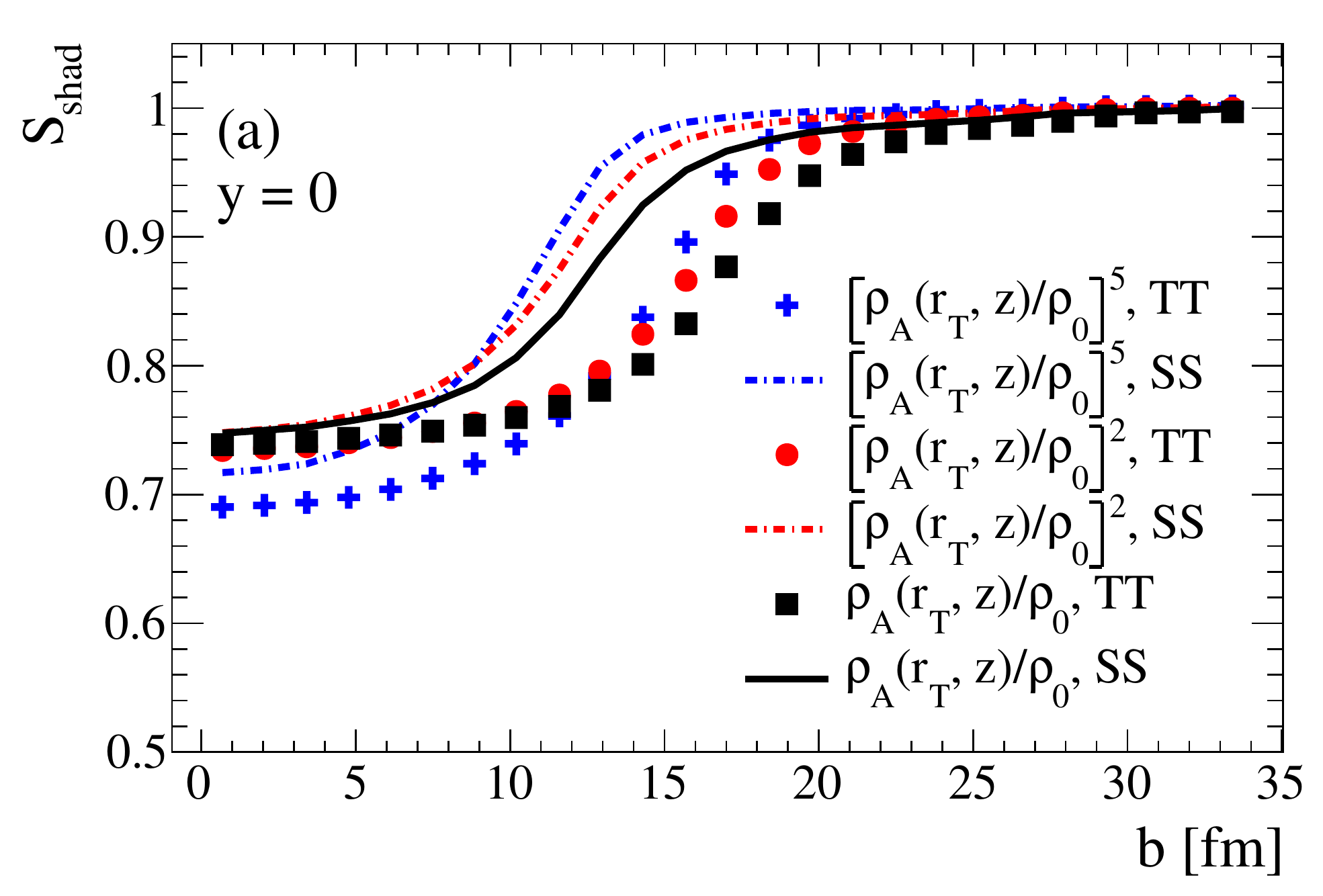} &
\includegraphics[width=0.48\textwidth]{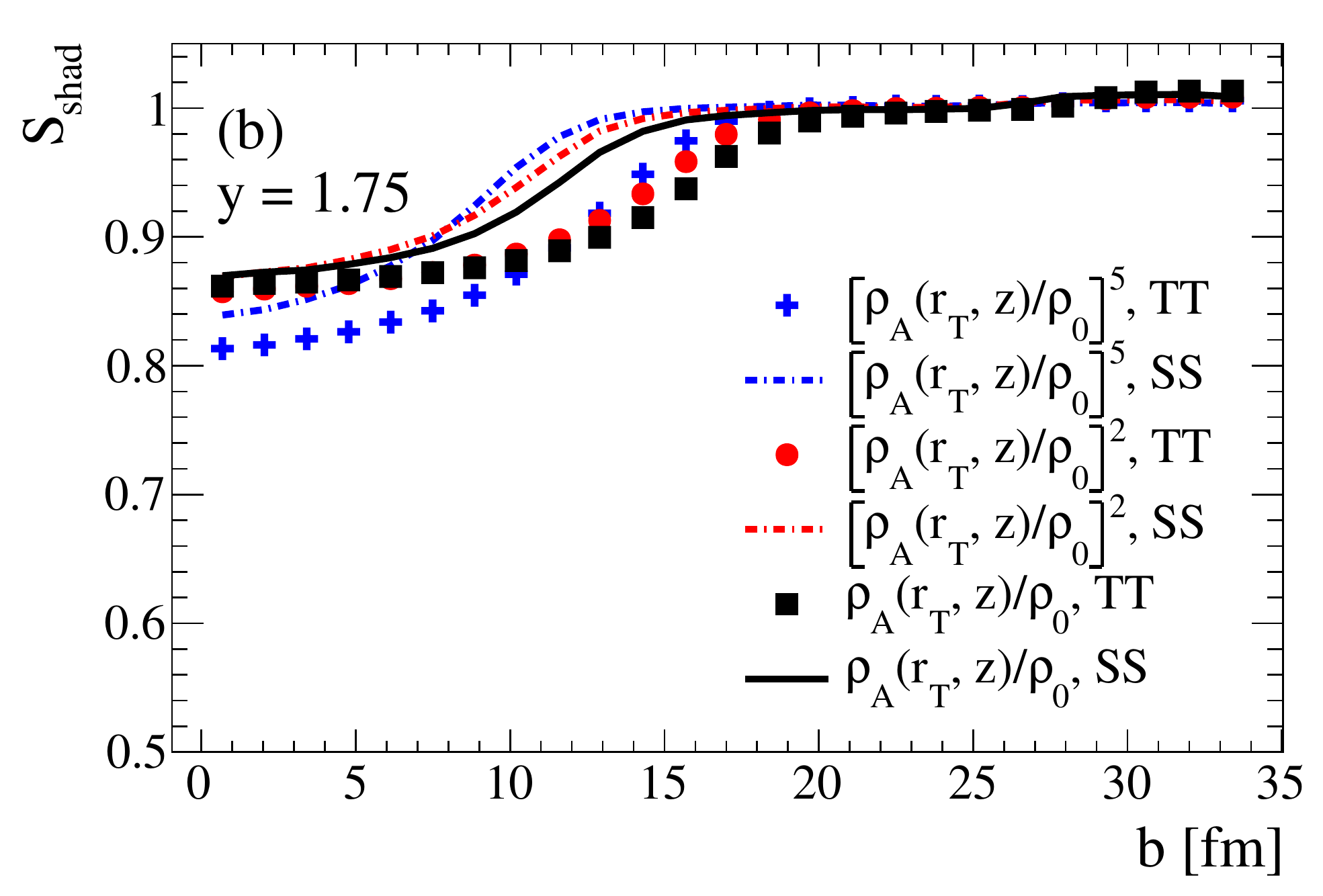} \\
\includegraphics[width=0.48\textwidth]{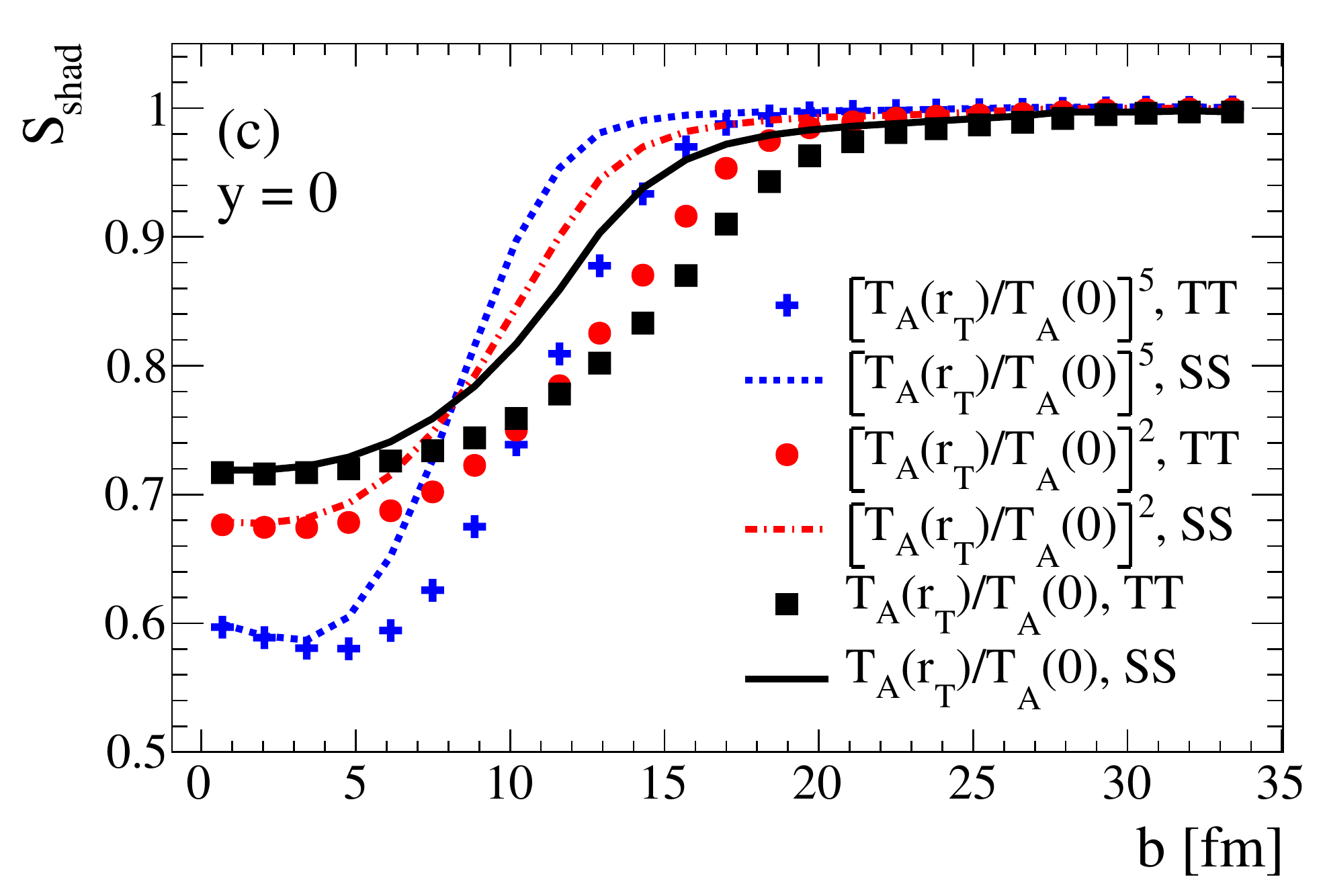} &
\includegraphics[width=0.48\textwidth]{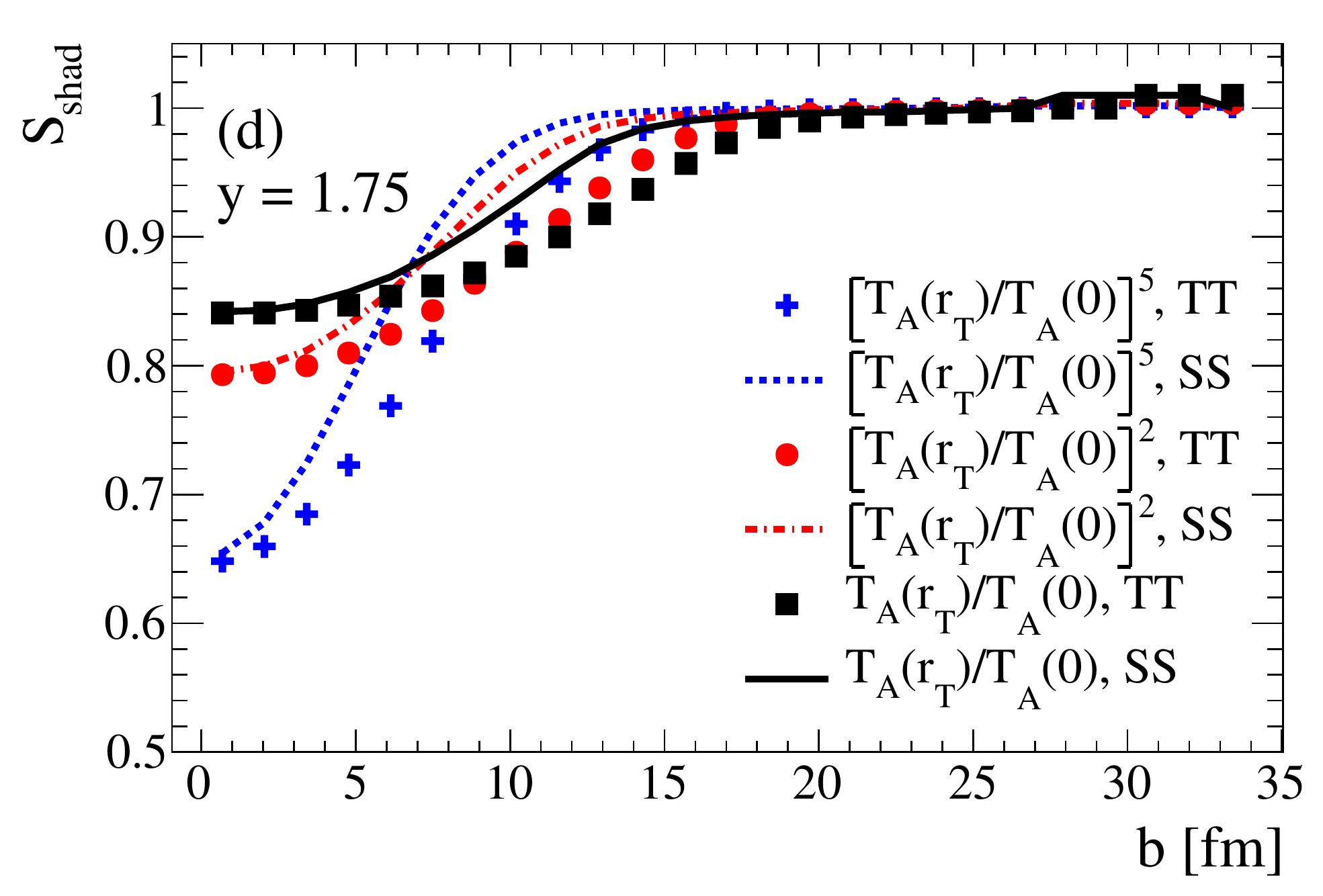} \\
\end{tabular}
\caption{\label{Fig:Jpsi_surv_probab_shadowing}(Color online) The 
survival probability due to shadowing, $S_{\rm shad}$ in TT 
(lines) and SS (symbols) orientations of U+U collisions for (a),(c) $y = 0$ 
and  (b),(d) $y=1.75$
assuming the impact parameter dependence of shadowing is proportional to $\rho_A$
(top) and $T_A$ (bottom) for $n = 1$, 2 and 5.}
\end{figure}

\begin{figure}[htdp]
\begin{tabular}{cc}
\includegraphics[width=0.48\textwidth]{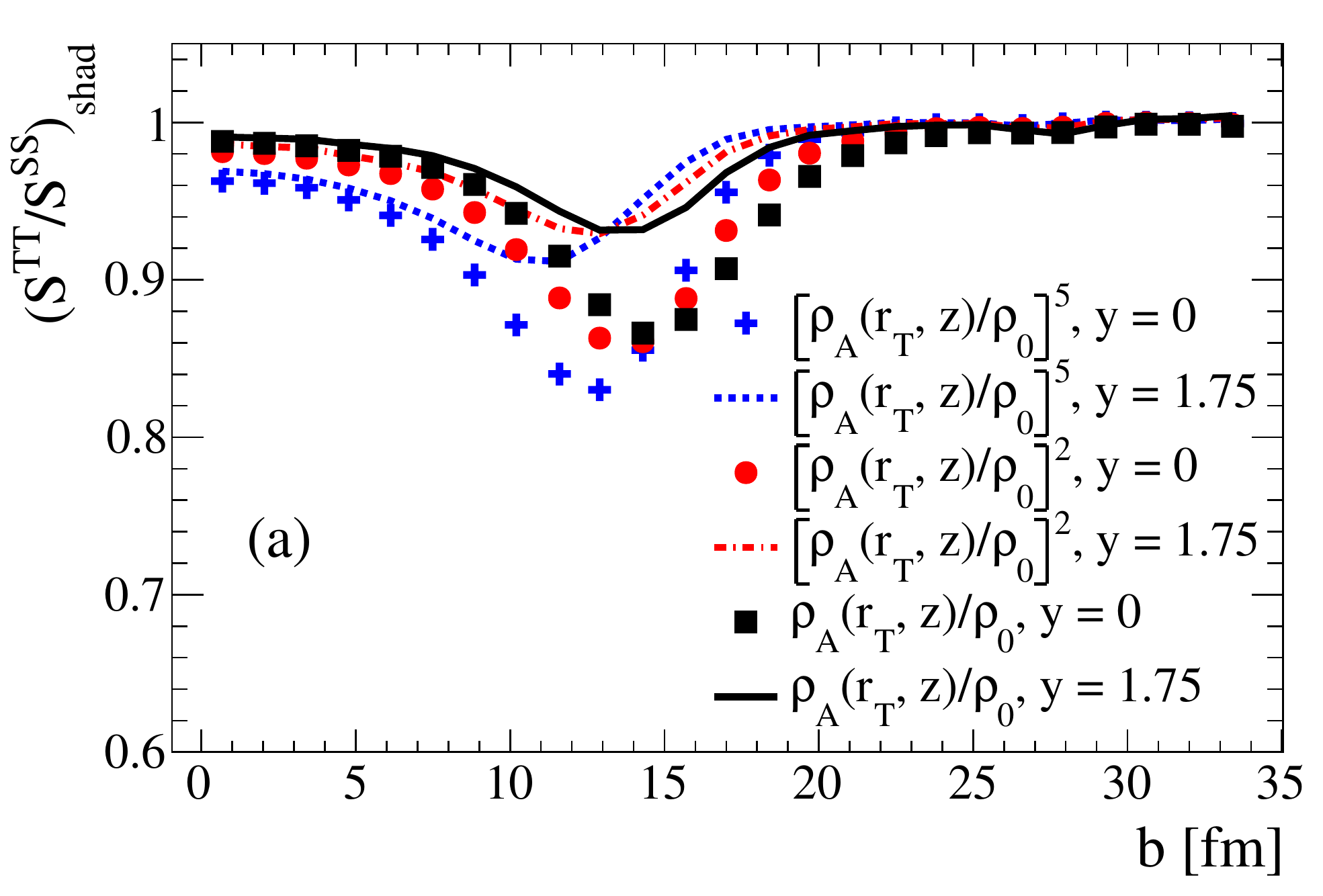} &
\includegraphics[width=0.48\textwidth]{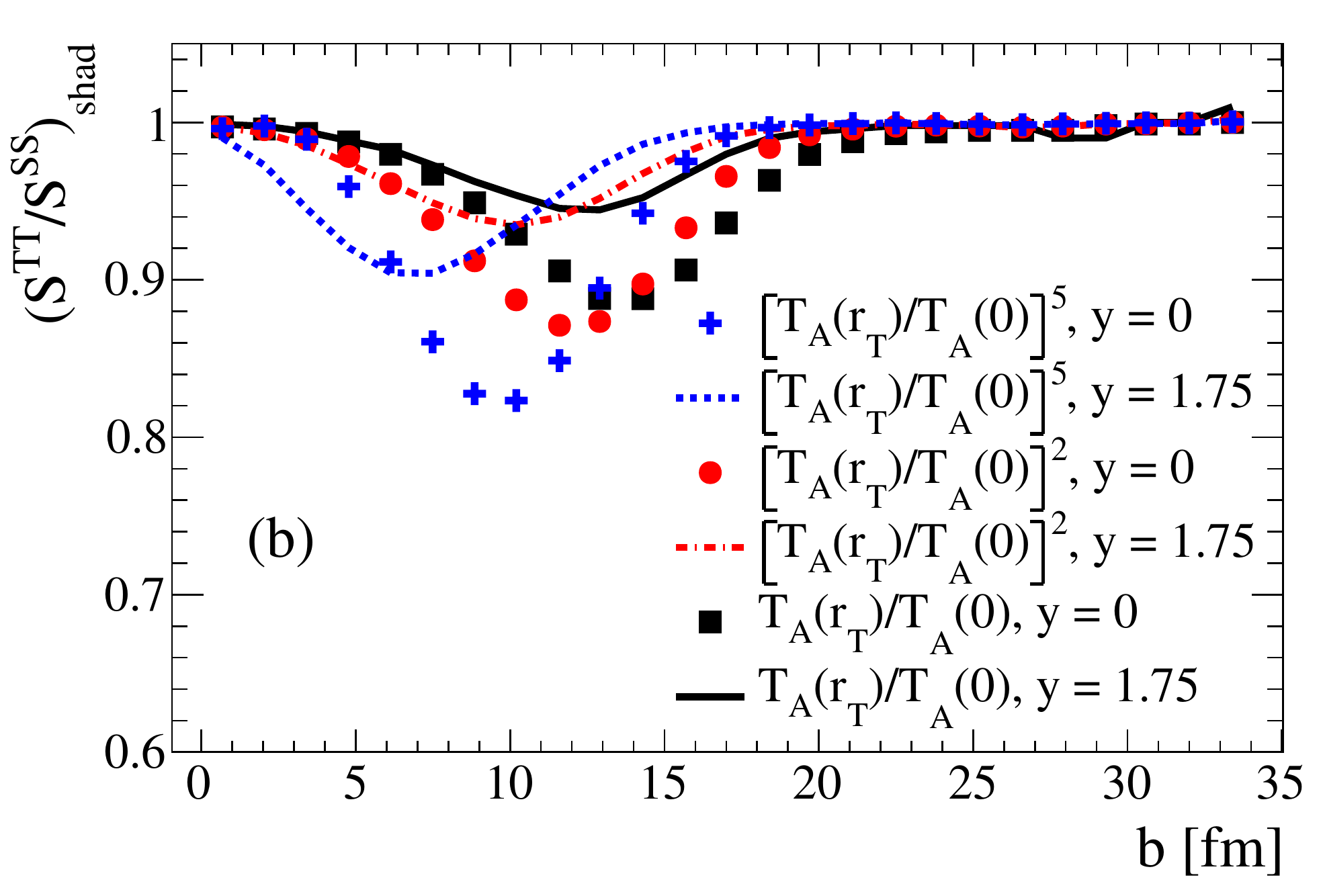} \\
\end{tabular}
\caption{\label{Fig:Jpsi_surv_probab_shadowing_ratio}(Color online) The ratio 
of charmonium survival probabilities, $S_{\rm shad}$, in TT and SS orientations 
assuming the impact parameter dependence of shadowing is proportional to $\rho_A$
(left) and $T_A$ (right) for $n = 1$, 2 and 5. }
\end{figure}

\begin{figure}[htdp]
\centering
\begin{tabular}{cc}
\includegraphics[width=0.48\textwidth]{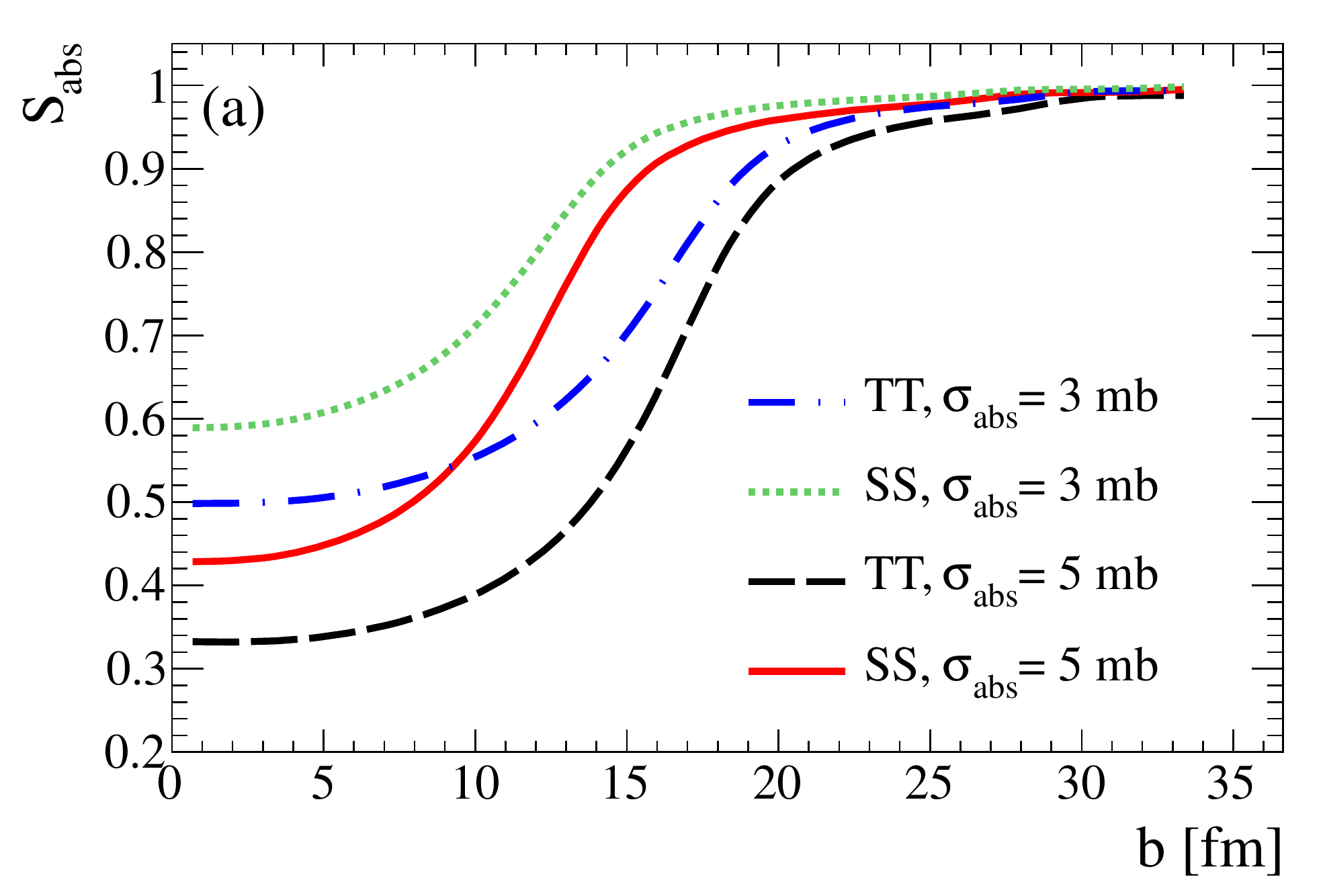} &
\includegraphics[width=0.48\textwidth]{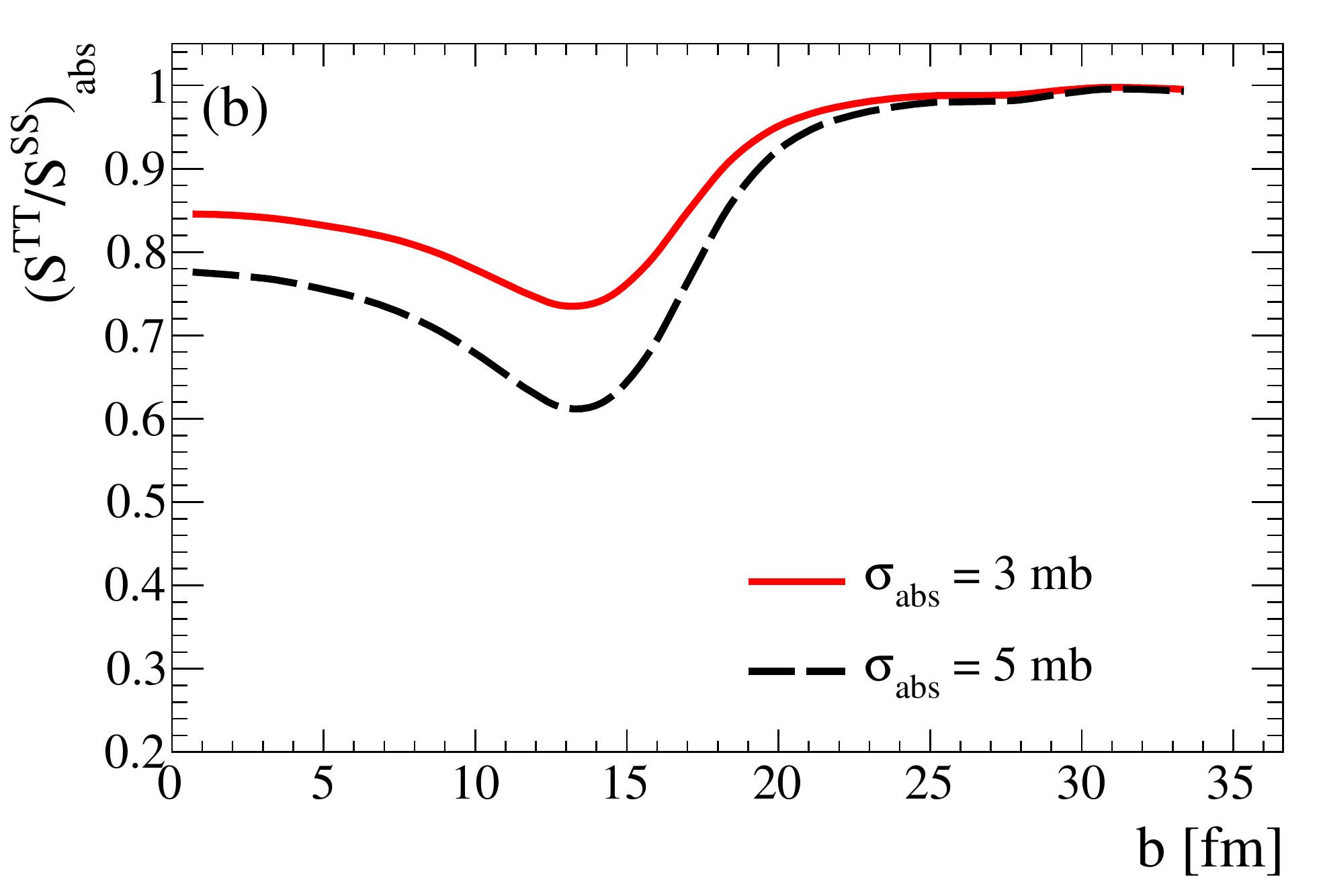} \\
\end{tabular}
\caption{\label{Fig:Jpsi_surv_probab_abs}(Color online) (a): The \jpsi\ 
nuclear absorption survival probability, \Sabs, in TT 
and SS orientations of U+U collisions for $\sigma_{\rm abs} = 3$ and 5
mb. (b): The ratio of results in TT and SS configurations for $\sigma_{\rm
abs} = 3$ (solid) and 5 (dashed) mb.}
\end{figure}

We begin with an investigation of the charmonium survival probability 
in U+U collisions for shadowing alone, \Sshad.
Figure~\ref{Fig:Jpsi_surv_probab_shadowing} shows \Sshad\ as a function of 
\bb\ in TT and SS configurations of 
U+U collisions at $y=0$ and 1.75, calculated in the color evaporation model
of quarkonium production \cite{RVTFTU}.  
(We do not show results at backward rapidity
since the results are symmetric around $y=0$ in $AA$ collisions.) 
Note that $S_{\rm shad} < 1$ represents 
shadowing, $S_{\rm shad} > 1$ antishadowing, and $S_{\rm shad} = 1$ no shadowing.
The convolution of two shadowing parameterizations results in stronger shadowing
at midrapidity than at forward rapidity for central EPS09 shadowing, as we
show in more detail later.  The CEM results exhibit a consistent pattern
between shadowing effects at leading and next-to-leading order \cite{RVgeneral}.
This behavior is typical of most shadowing
parameterizations based on collinear factorization \cite{rvhip}.  

Calculations of color singlet $J/\psi$ interactions in the 
dipole approximation suggest that collinear
factorization may be inapplicable due to the coherence of the interaction
\cite{tuchin,boris}.   In this case, higher-twist effects enhanced by powers 
of $A^{1/3}$ would dominate $pA$ interactions and enhanced $J/\psi$
suppression should set in at large rapidity.  However, such
enhanced effects are seen in fixed-target energies as well, outside the range
of validity of gluon saturation models \cite{CarlosHermineme,QWGdoc}, and
appear to scale with projectile momentum fractions, $x_1$, rather than with
the target fraction $x_2$.  

While the most recent PHENIX data
\cite{Adare:2010fn} seem to exhibit a stronger than linear impact parameter 
dependence of shadowing \cite{Jamie}, the min bias results for $R_{\rm dAu}$
are in good agreement with calculations based on Ref.~\cite{RVgeneral}, as are
the results in the most central impact parameter bin.  Indeed, these new
data suggest that while the shadowing may decrease more strongly than the 
calculations of Refs.~\cite{SpatialShadDep1999,Inhomogeneous:Shadowing}
predict, the strength increases rather slowly for low impact parameters.

To illustrate the effect of a larger than linear impact parameter dependence of
shadowing, we compare the results for shadowing parameterizations that depend
on the local nuclear matter density and on the parton path length through 
nuclear matter
\cite{Inhomogeneous:Shadowing},
\begin{eqnarray} 
S^i_{\rho} (A,x,Q^2,\vec{r}_{T},z) & = & 1 + N_{\rho} (S^i(A,x,Q^2) - 1) 
\bigg(\frac{\rho_A(\vec{r}_{T},z)}{\rho_0}\bigg)^n \label{rhoparam} \, \, , \\
S^i_{T_A} (A,x,Q^2,\vec{r}_{T},z) & = & 1 + N_{T_A} (S^i(A,x,Q^2) - 1) 
\bigg(\frac{\int dz
\rho_A(\vec{r}_{T},z)}{\int dz \rho_A(0,z)}\bigg)^n \nonumber \, \, , \\
                  & = & 1 + N_{T_A} (S^i(A,x,Q^2) - 1) 
\bigg(\frac{T_A(\vec{r}_{T})}{T_A(0)}\bigg)^n \label{TAparam} \, \, ,
\end{eqnarray}
where $T_A$ is the nuclear profile function and
$N_{\rho}$, $N_{T_A}$ are chosen to normalize the integral of
$S^i(A,x,Q^2,\vec{r}_{T},z)$ over the nuclear volume, weighted by the nuclear 
density distribution, to $S^i(A,x,Q^2)$ in minbias collisions.  The values of the normalizations
are dependent on the power of $n$ chosen as well as the parameterization of
the impact parameter dependence.  Results are shown in 
Fig.~\ref{Fig:Jpsi_surv_probab_shadowing} for $n = 1$, 2 and 5. 

When the shadowing is parameterized according to the nuclear density, as
in Eq.~(\ref{rhoparam}), the difference between the TT and SS configurations is
rather distinct over a broad range of impact parameters.
The shape of $S_{\rm shad}(b)$ resembles the Woods-Saxon density distribution. 
The normalization, $N_{\rho}$, changes slowly with $n$.  The difference
between the two spatial orientations of U+U collisions, TT and SS, is very
small for most central collisions, $b < 6$ fm.  As $n$ increases, $N_\rho$
increases, as does the difference between the TT and SS orientations.  
Nonetheless, the results retain the general Woods-Saxon shape while impact
parameter dependence steepens over the transition from shadowing to free 
nucleon behavior.  This transition is centered around $b\sim 2R_C$ for $n=1$, the
impact parameter where the effect of shadowing is half the $b=0$ value.
Note, that $R_C = R_{\rm T}$, 
the length of the long axis in TT configurations, and $R_C = R_{\rm S}$ in SS
configurations.   

When the longitudinal direction is integrated over, as
in Eq.~(\ref{TAparam}), the distinctive Woods-Saxon shape seen in the upper
half of Fig.~\ref{Fig:Jpsi_surv_probab_shadowing} is somewhat
washed out and $S_{\rm shad}$ increases rather smoothly with impact parameter.  
A separation between the TT and SS configurations is already apparent for
small values of $b$.  In this case, $N_{T_A}$ depends more strongly on $n$
so that the difference between $n = 1$ and $n=2$ is larger here than for
Eq.~(\ref{rhoparam}).

Increasing $n$ reduces the effective radius for
shadowing in both formulations in Eqs.~(\ref{rhoparam}) and (\ref{TAparam}).
We have only shown results for $n \leq 5$ but have checked higher values of
$n$ and see that the trend continues.  The separation between the TT and SS
configurations for Eq.~(\ref{rhoparam}) increases to 10\% at low $b$ for $n=15$
while the impact parameter dependence of $S_{\rm shad}$ more closely resembles
a step function.  The growth of $N_{\rho}$ with $n$ remains slow.  On the
other hand, with the dependence of Eq.~(\ref{TAparam}), the difference between
the orientations remains small, even at higher $n$, while $N_{T_A}$ is a stronger
function of $n$.  Large values of $n$ are still consistent with the 
PHENIX data at forward rapidity \cite{RVDMTF}. 

We expect
that the impact parameter dependence can be characterized by other means 
and only show the possible range of effects here to
study the sensitivity of CNM effects in U+U collisions on the impact parameter
dependence of shadowing.  A full analysis of the $b$ dependence,
under study in Ref.~\cite{RVDMTF}, is beyond 
the scope of this work.

The ratio of \Sshad\ in the two configurations is shown in 
Fig.~\ref{Fig:Jpsi_surv_probab_shadowing_ratio} for the same values
of $y$ and $n$.  Away from the
transition region, the ratio is relatively constant and close to unity with a
maximum 5\% difference employing 
Eq.~(\ref{rhoparam}) with $n=5$.  The ratio of results with
Eq.~(\ref{TAparam}) is much smaller at $b \sim 0$.  Thus, for moderate 
shadowing, the ratio in central collisions is relatively
insensitive to $S_{\rm shad}$. The largest change in the ratio with $n=1$
is at $\sim (R_{\rm T} + R_{\rm S})/2$.  As $n$ increases, the dip in the ratio
TT/SS deepens and shifts to lower impact parameters.  The dip is deeper for
$y=0$ where the midrapidity shadowing effect is stronger.  The forward 
rapidity effect is weaker but tends to shift to still lower impact parameters
because the assumption of factorization pairs strong shadowing in one nucleus
with small shadowing or some antishadowing in the other.
In both cases, since $f_i^A(x,Q^2,\vec{r}_{T},z)$ is
proportional to the nuclear density distribution, see Eq.~(\ref{fanuc}), the
shape of \Sshad$(b)$ directly reflects the spatial orientations of the U nuclei.

We now discuss the impact parameter dependence of the survival probability for
nuclear absorption.  Figure~\ref{Fig:Jpsi_surv_probab_abs} (a)
shows \Sabs\ as a function of \bb\ for U+U collisions in the TT and SS 
configurations assuming $\sigma_{\rm abs} = 3$ and $5$~mb. These two values of 
\SigAbs\ bracket the range obtained by the RHIC experiments. As previously discussed, the min bias 
result, $R_{\rm dAu}$, reported recently by PHENIX~\cite{Adare:2010fn} is 
relatively well described by the EPS09 shadowing parameterization with 
$\sigma_{\rm abs} = 4$ mb. Before the most recent PHENIX data \cite{Adare:2010fn}
were available, the divergence of $R_{CP}(y)$ from the calculations of absorption
with shadowing was quantified the extraction of a rapidity-dependent absorption
cross section \cite{QWGdoc}.  These results showed a strong increase of the
effective absorption cross section at forward rapidity which could be
attributed to a heretofore neglected effect such as initial-state energy loss.
However, a reanalysis of these data based on a more complete 
understanding of the impact parameter dependence of nuclear shadowing would
affect the magnitude of the extracted absorption cross section.  This will be
addressed in Ref.~\cite{RVDMTF}.

While a similar trend as a function of $b$ is observed relative to Fig.~\ref{Fig:Jpsi_surv_probab_shadowing}, the
 difference between TT and SS orientations at $b=0$ is larger for absorption.  Thus
 nuclear absorption is the only relevant orientation-dependent
 CNM effect in central collisions.  The larger effect is due to the exponential factor in \Sabs, proportional to the path
 length, in addition to the overall dependence on the density in the calculation of the total
 $J/\psi$ yield. Thus the angular orientation of the colliding nuclei affects
 absorption more strongly than shadowing. The exponential dependence on
 density in \Sabs\ also broadens the transition region around $b \sim 2R_C$.

The ratio of survival probabilities for the two orientations is shown in
Fig.~\ref{Fig:Jpsi_surv_probab_abs}(b) for $\sigma_{\rm abs} = 3$ 
and 5 mb. The ratio shows a $\sim 20$\% difference relative to the maximum
5\% for shadowing alone.  It also varies more slowly with impact 
parameter, as seen by comparing 
Figs.~\ref{Fig:Jpsi_surv_probab_shadowing_ratio} and 
\ref{Fig:Jpsi_surv_probab_abs}(b).

\begin{figure}[htdp]
\begin{tabular}{cc}
\includegraphics[width=0.48\textwidth]{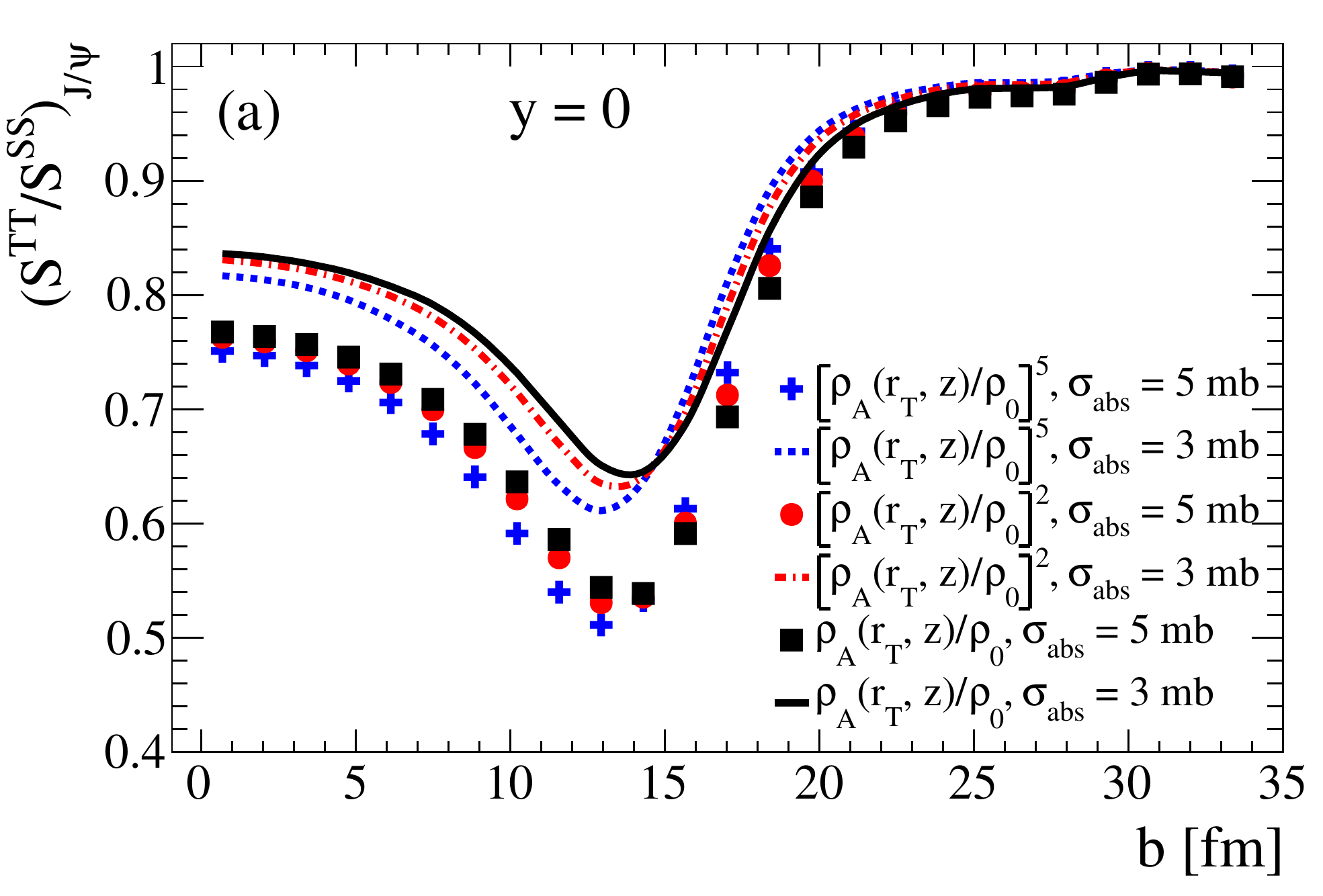} &
\includegraphics[width=0.48\textwidth]{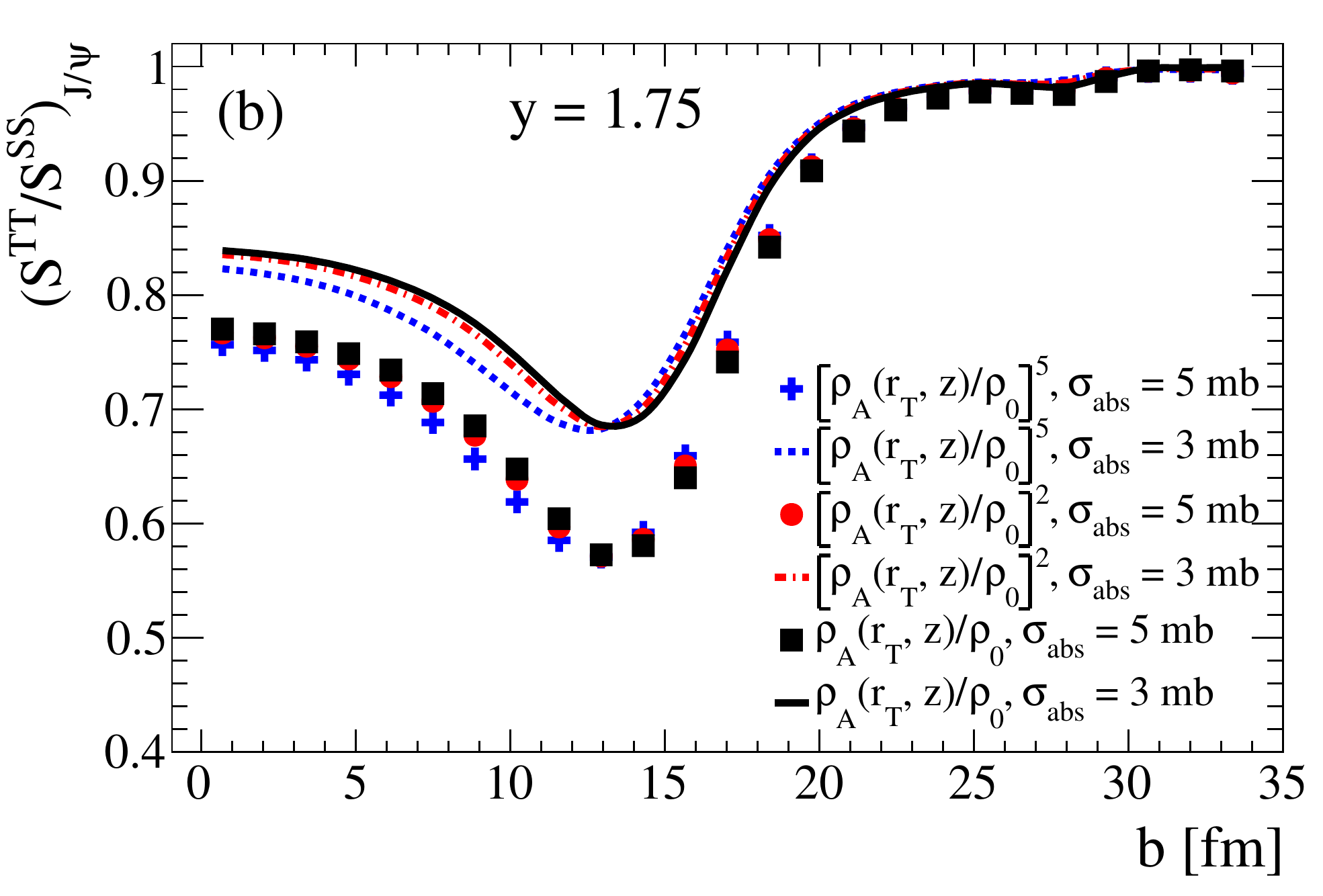} \\ 
\includegraphics[width=0.48\textwidth]{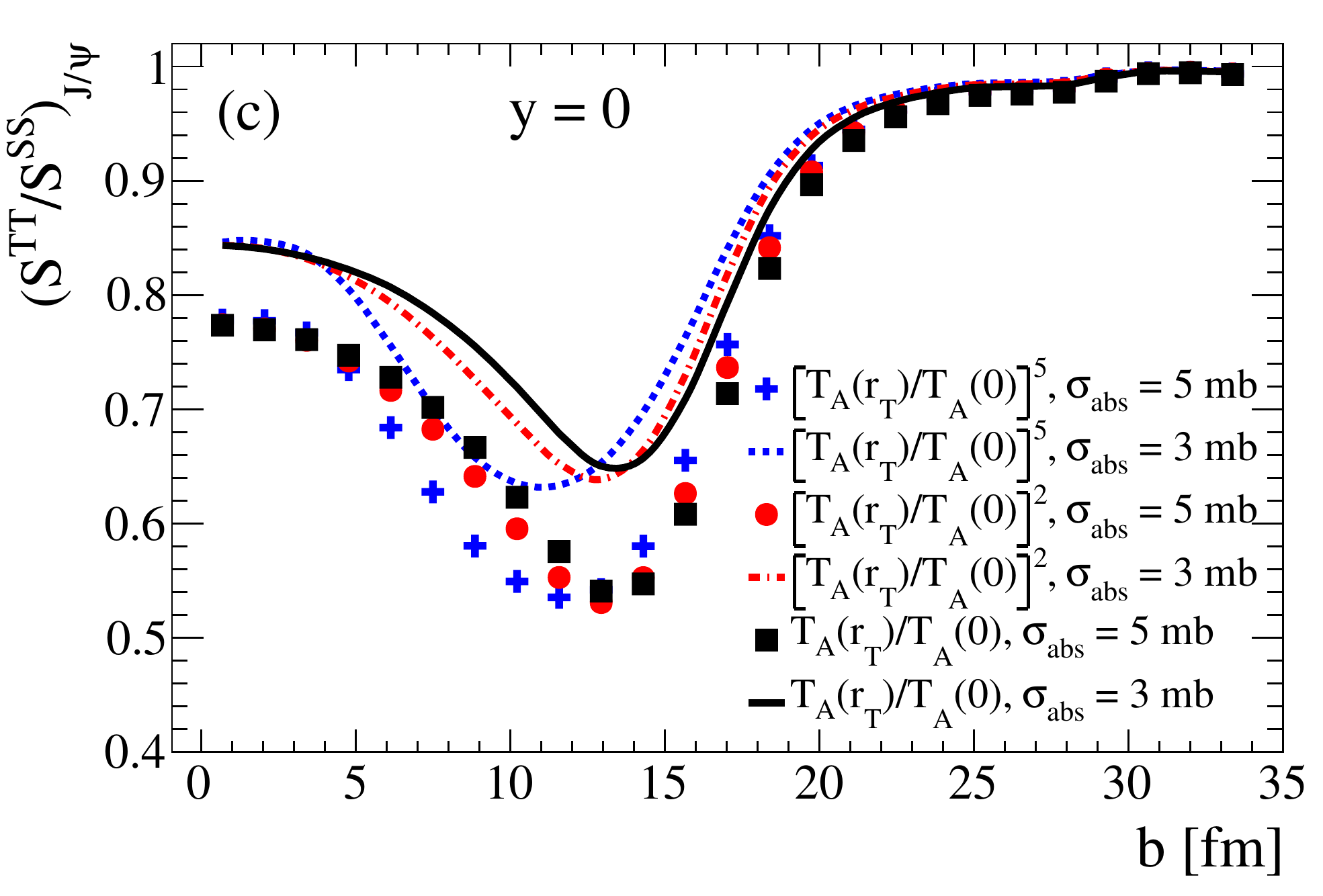} &
\includegraphics[width=0.48\textwidth]{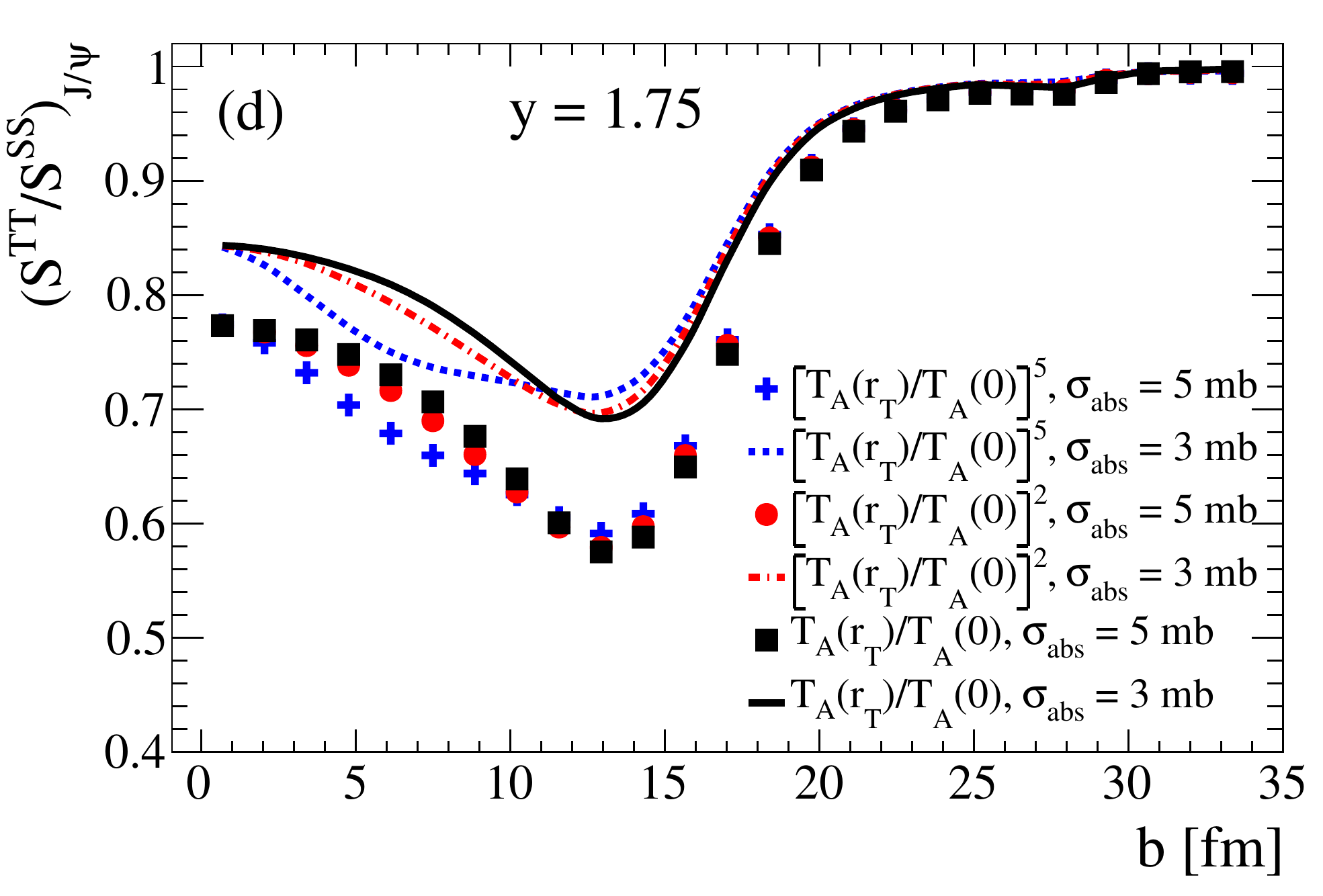} \\ 
\end{tabular}
\caption{\label{Fig:Overal_Jpsi_surv_probab_ratio}(Color online) The ratio of the cold
nuclear matter \jpsi\ survival probability in TT relative to SS orientations 
of U+U collisions for \SigAbs\ =  3 and 5 mb assuming different parameterizations of the impact parameter dependence of shadowing
(see text for details).}
\end{figure}

Figure~\ref{Fig:Overal_Jpsi_surv_probab_ratio} shows the ratios of the total
\jpsi\ survival probability in the TT and SS configurations
due to cold nuclear matter effects, $S_{J/\psi} = 
S_{\textrm{abs}} S_{\textrm{shad}}$, as a function of impact parameter. 
In central collisions, $b<5$~fm, the magnitude of \SpsiTtSs\ is most 
sensitive to \SigAbs\ while at larger impact parameters, $b \geq 10$~fm,
\SpsiTtSs\ is dominated by shadowing. 
If the TT and SS orientations could be effectively selected in an experiment, 
these features could help differentiate cold nuclear matter effects from
color screening effects.  However, the best case scenario would be to first
attempt to select the tip and side orientations of the uranium nuclei in d+U
collisions.

\begin{figure}[htdp]
\begin{tabular}{cc}
\includegraphics[width=0.48\textwidth]{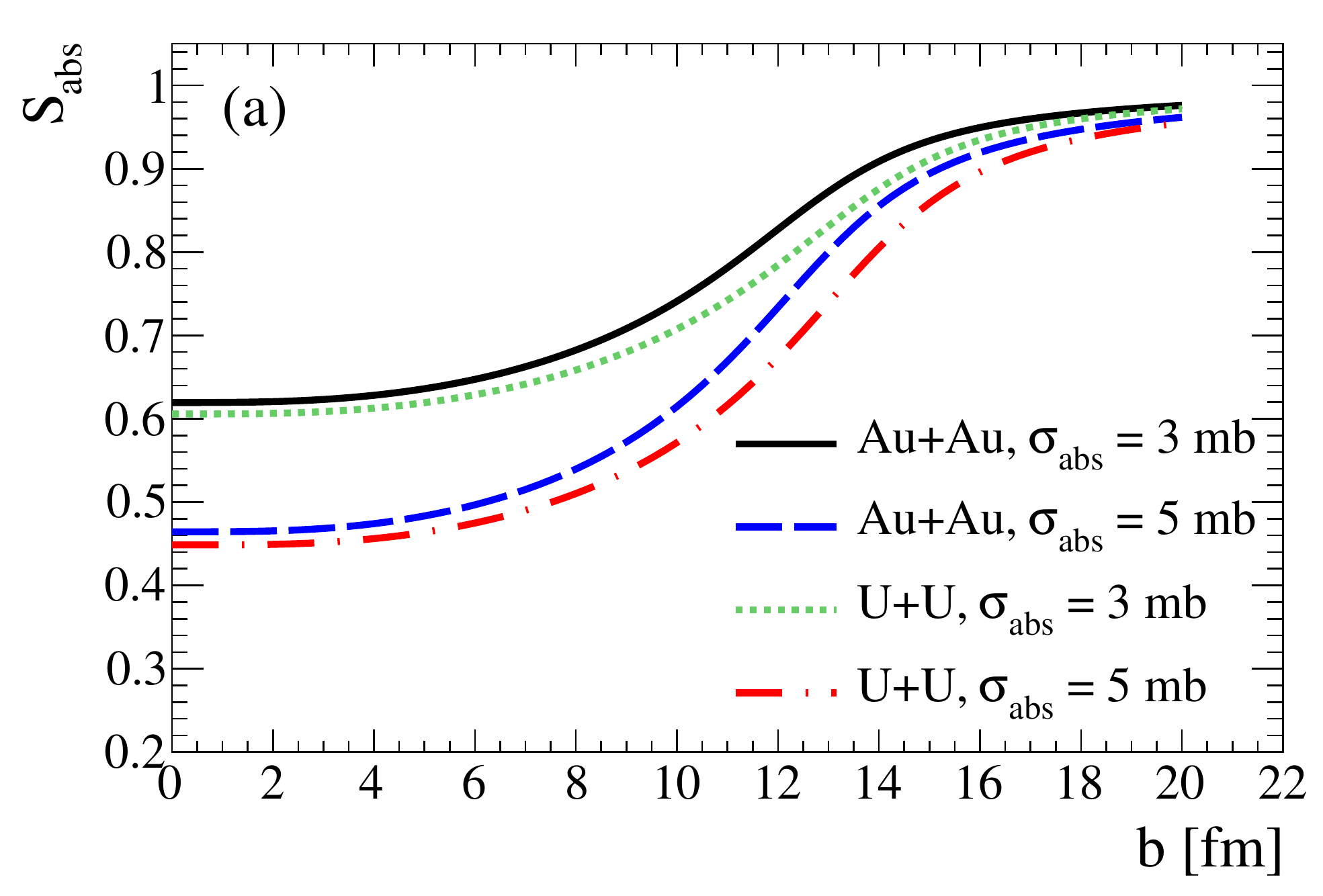} &
\includegraphics[width=0.48\textwidth]{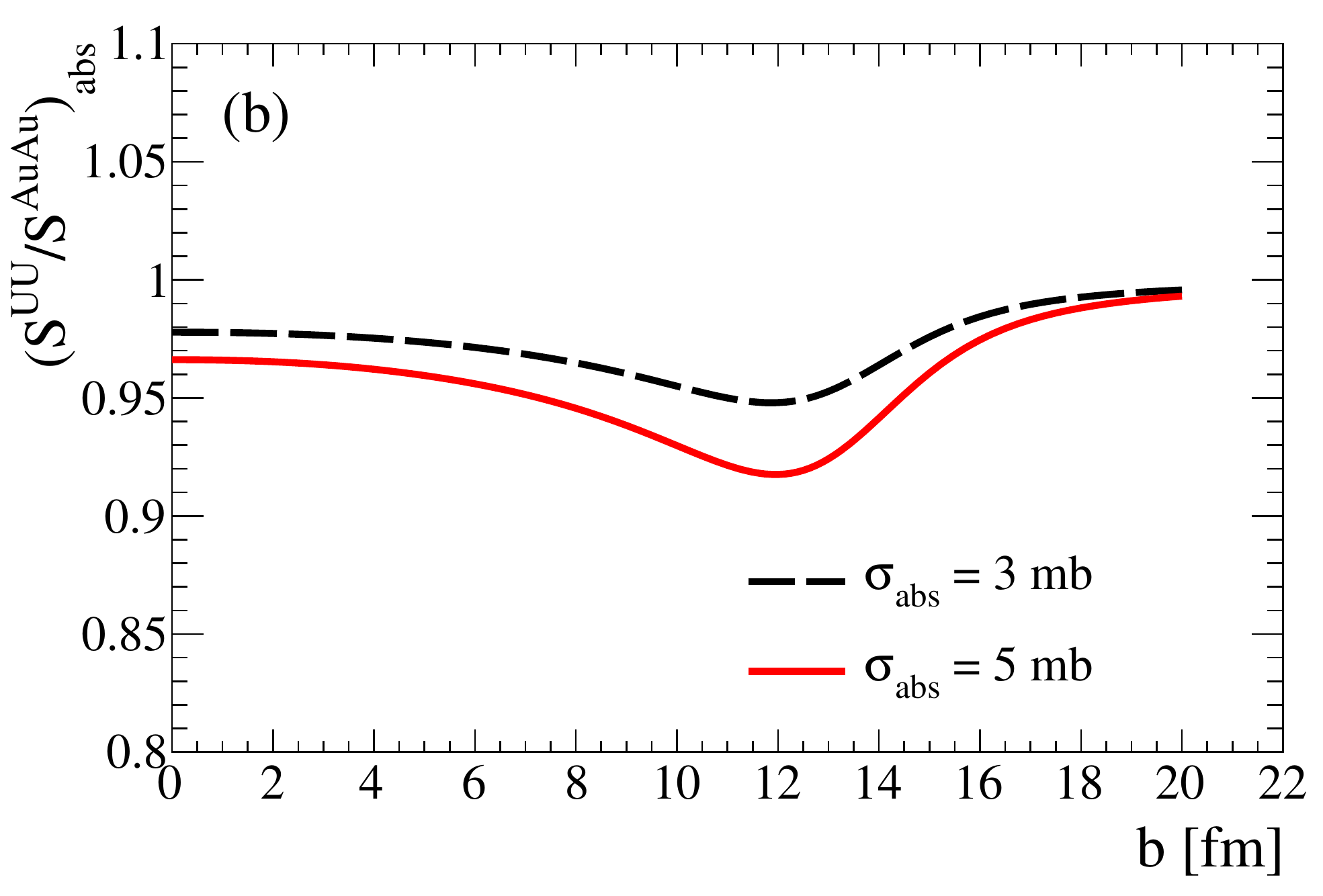} \\
\end{tabular}
\caption{\label{Fig:Jpsi_surv_probab_UU_AuAu}(Color online) Nuclear absorption as a function 
of $b$ for Au+Au and U+U collisions averaged over all orientations 
(a) and the ratio of U+U relative Au+Au collisions (b).}
\end{figure}

Even if the TT or SS configurations cannot be selected due to either technical 
difficulties or low efficiencies, U+U collisions averaged over all orientations 
could be used for quarkonium studies. There are no significant experimental 
difficulties and the nuclear path length $L$ and charged track density are 
respectively 5-10\% and 10-15\% larger for the same centrality class in 
Au+Au collisions \cite{Masui2009440}. The larger $L$ increases the effective 
nuclear absorption. Figure~\ref{Fig:Jpsi_surv_probab_UU_AuAu} shows the
orientation-integrated \Sabs\ as a function of impact parameter for Au+Au 
and U+U collisions with \SigAbs\ = 3 and 5 mb. The difference between 
$S_{\textrm{abs}}^{\textrm{UU}}$ and $S_{\textrm{abs}}^{\textrm{AuAu}}$ is largest, $\sim 5$\%, 
at  $b\sim 11-12$~fm and can be attributed to the difference in radii, 6.81~fm 
for uranium and 6.38~fm for gold.
The magnitude of the difference depends on \SigAbs.  

\begin{figure}[htdp]
\begin{tabular}{cc}
\includegraphics[width=0.48\textwidth]{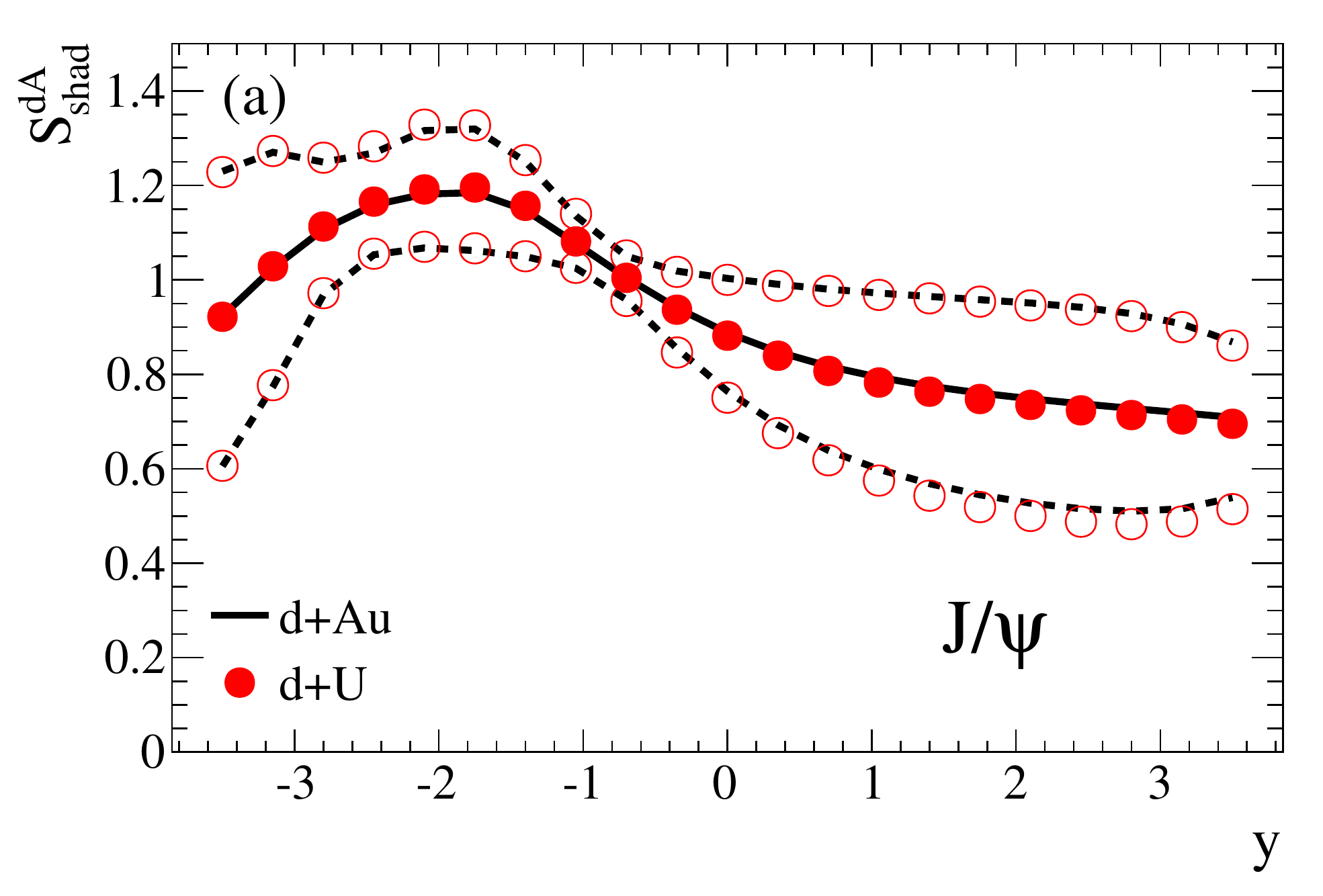} &
\includegraphics[width=0.48\textwidth]{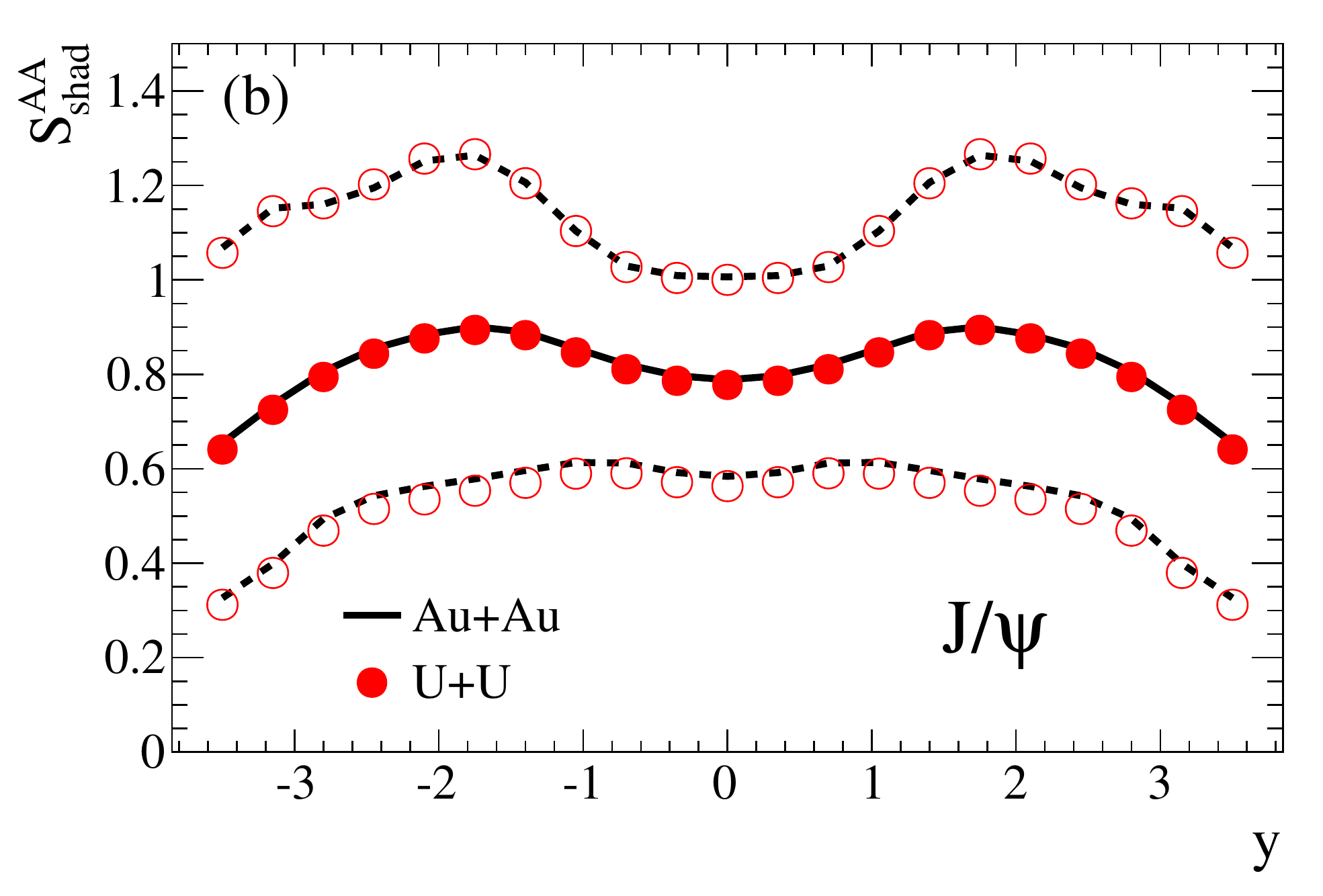} \\ 
\end{tabular}
\caption{\label{Fig:Jpsi_RdAu}(Color online) The \jpsi\ survival probability  due to 
shadowing in d+$A$ (a) and $A+A$ (b) collisions. The solid lines (filled circles) 
represent results for Au (U) nuclei obtained with the central EPS09
shadowing parametrization while the dashed lines (open circles) show the
results with the upper and lower limits of the EPS09 uncertainty.}
\end{figure}

On the other hand, the relative difference in the shadowing effect is 
negligible. The expected \Sshad\ is shown in Fig.~\ref{Fig:Jpsi_RdAu} for 
d+Au and d+U collisions (a) as well as Au+Au and U+U collisions (b). 
Shadowing effects in U+U and Au+Au collisions at the same \sNN are very 
similar at a given value of \bb.  Thus the uncertainty due to shadowing 
will cancel in the ratio of the \jpsi\ yields in U+U and Au+Au 
collisions.

\begin{figure}[htdp]
\begin{tabular}{cc}
\includegraphics[width=0.48\textwidth]{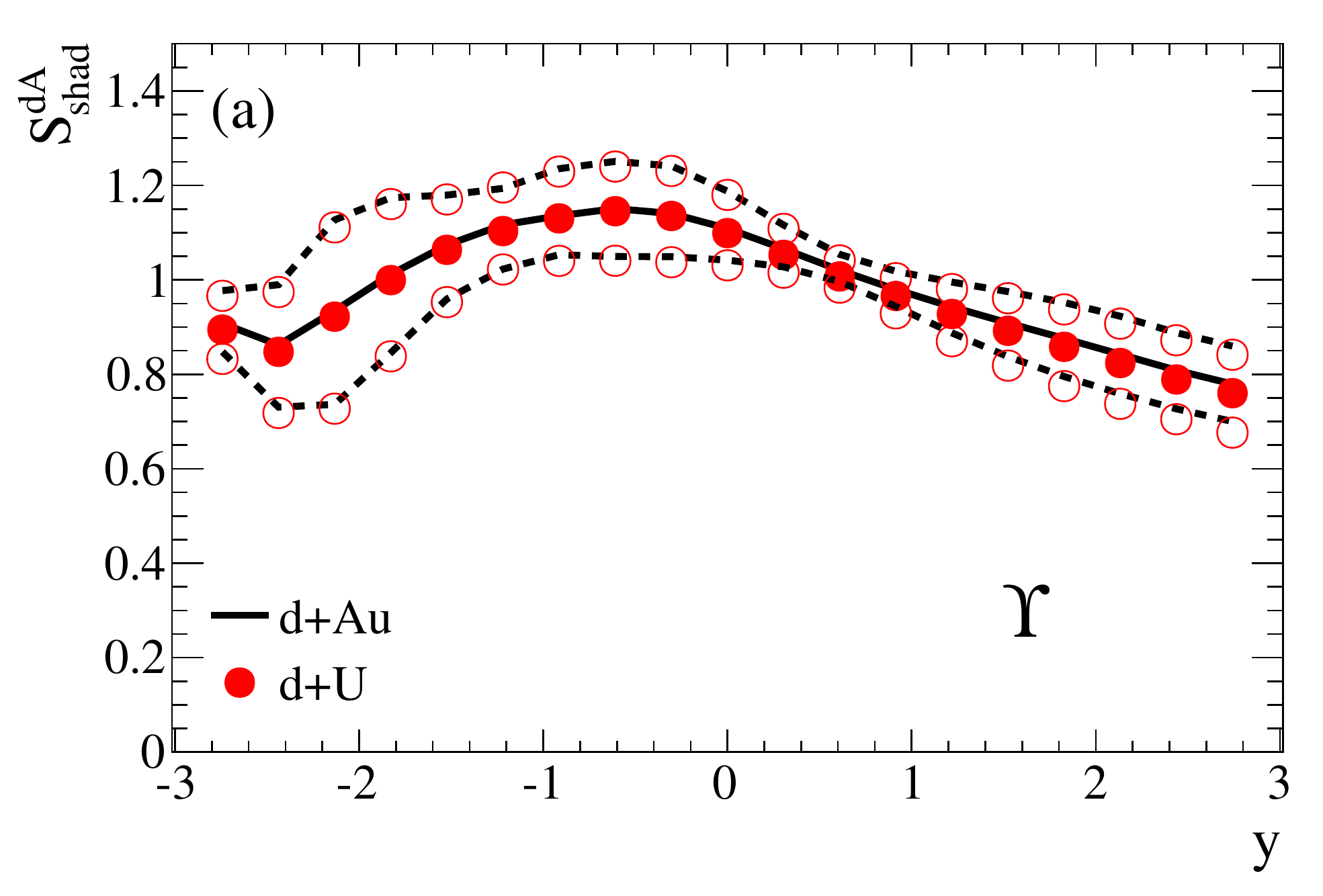} &
\includegraphics[width=0.48\textwidth]{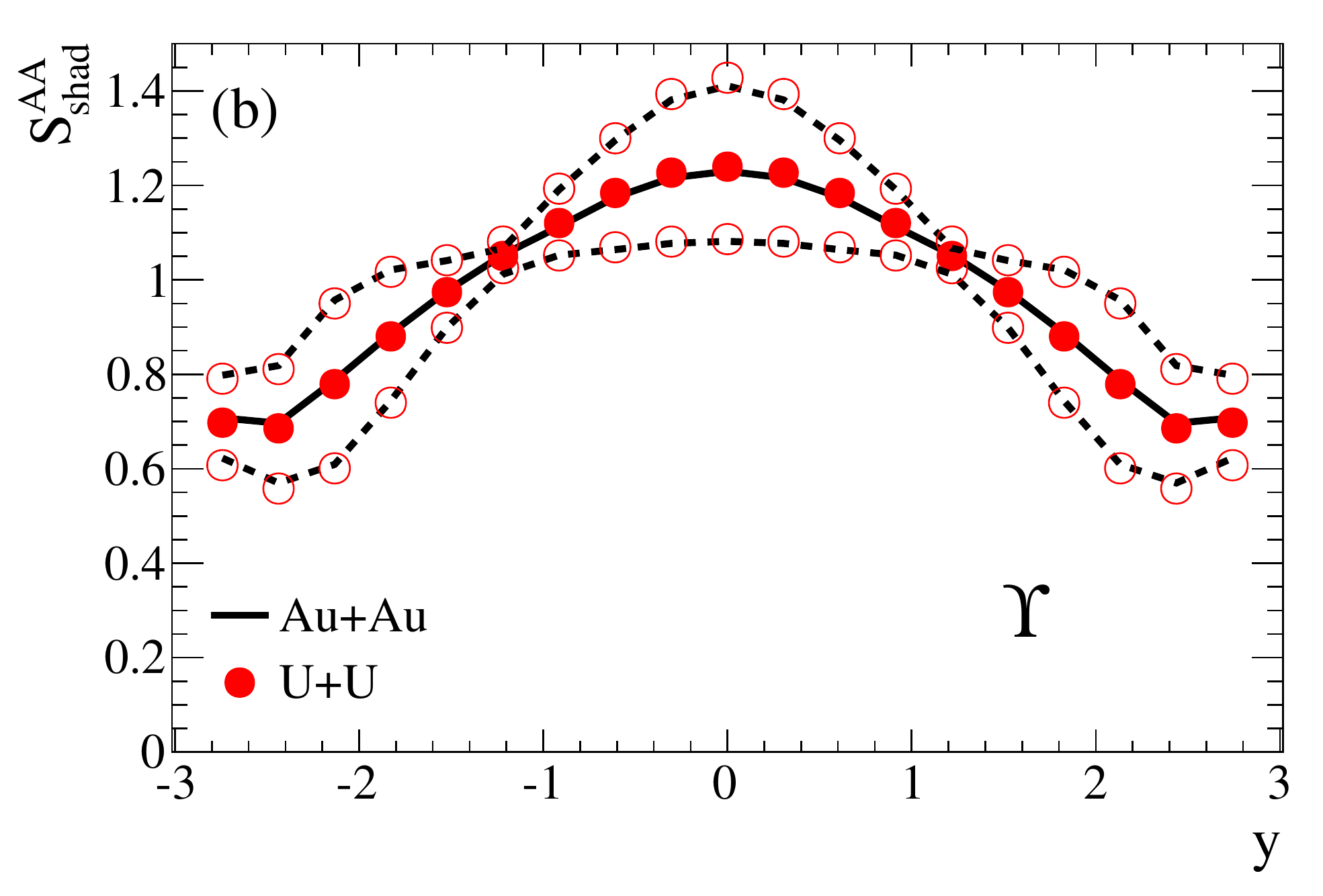} \\ 
\end{tabular}
\caption{\label{Fig:Ups_RdAu}(Color online) The \Ups\ survival probability  due to 
shadowing in d+$A$ (a) and $A+A$ (b) collisions. The solid lines (filled circles) 
represent results for Au (U) nuclei obtained with the central EPS09
shadowing parametrization while the dashed lines (open circles) show the
results with the upper and lower limits of the EPS09 uncertainty.}
\end{figure}

We now discuss the modification of \Ups\ production in Au+Au and U+U collisions 
due to cold nuclear matter effects.  The nuclear absorption of \Ups$(1S)$ 
seems to be 
significantly weaker than for \jpsi\ \cite{Vogt:2010aa}. Moreover, antishadowing
may be expected at midrapidity rather than shadowing, supported by the 
experimental results at RHIC.  The cross section of \Ups\ production in d+Au 
collisions is well described by the color evaporation model with antishadowing 
and no nuclear absorption \cite{Upsilon:dAu}. The expected \Sshad\ for \Ups\ 
is shown in Fig.~\ref{Fig:Ups_RdAu} for d+Au and d+U collisions (a) as well 
as for Au+Au and U+U collisions (b). The effect of shadowing is much 
smaller than for \jpsi.  Instead, antishadowing at midrapidity may increase 
the \Ups\ 
rate by as much as 20-30\%. Note that antishadowing effects in U+U and Au+Au 
collisions at the same \sNN\ are very similar, as is the case for the \jpsi.
Furthermore, a different formulation of the quarkonium production model,
such as in Ref.~\cite{lansberg}, would yield similar results even though the
larger average scale used in those calculations further reduces the overall
shadowing effect.

Finally, we comment on the suggestion that double-color filtering and mutual 
boosting of the saturation scales in colliding nuclei make the transition of 
cold nuclear matter effects from $pA$ to $AA$ collisions nontrivial 
\cite{boris}.  However these effects are rather small at RHIC energies (less 
than 10\% for double-color filtering and 18\% for
mutual boosting of the saturation scales). In addition, they act in the opposite
directions, therefore canceling to a large extent.  Reference~\cite{boris} 
proposed that an increase of \jpsi\ $p_T$ broadening in $AA$ collisions relative
to $pA$ would directly reflect the
boosting effect. Such an increase is not seen at RHIC.  In fact, 
$\langle p_T^2 \rangle$ at midrapidity in Au+Au collisions at \sNN$ =200$ GeV 
is approximately independent of centrality  \cite{Adare:2008sh}.
Determining the nature and strength of such a boosting effect at RHIC
requires high precision data.  Moreover, the effects proposed in 
Ref.~\cite{boris} will be very similar in Au+Au and U+U
collisions and thus cancel if ratios of $J/\psi$ production in the two
systems are studied.

\section{Energy density and temperature in U+U collisions}

We now investigate what energy densities and temperatures are accessible in 
U+U collisions. The charged track density is expected to increase in U+U 
interactions compared to Au+Au.  Therefore, the U+U temperature should be
higher in the deconfined phase.  Employing U+U collisions could extend the 
range of temperatures accessible at RHIC and test the sequential melting
hypothesis of charmonium and bottomonium states. Suppression of $\Upsilon(1S)$ 
beyond CNM effects would constitute direct proof of color screening since 
secondary $\Upsilon$ production by coalescence from a QGP is negligible at
RHIC.  Comover absorption is also insignificant.  Recent STAR results show that 
the $\Upsilon(1S)$ is not strongly suppressed in central Au+Au collisions 
\cite{STAR:Upsilon:HP2010}. Below, we investigate whether the temperature of
the matter produced in U+U collisions could be high enough to expect 
``melting" of the $\Upsilon(1S)$ state in addition to $J/\psi$ suppression. 

We use the Bjorken formula \cite{Bjorken:1982qr} to estimate the energy density 
available in U+U collisions. The Bjorken energy density in a one-dimensional 
longitudinally expanding system can be calculated using
\begin{equation}
\epsilon_{\textrm{B}} = \frac{1}{\tau S_\perp} \frac{dE_{\textrm{T}}}{dy}
\end{equation}
where $\tau$ is the formation time of the medium, $S_\perp$ is the transverse 
area overlap of the colliding nuclei, and \dEdy\ is the transverse energy 
density. We estimate \dEdy\ from PHENIX data \cite{Adler:2004zn},
\begin{equation}
\frac{dE_{T}}{dy} = C_{\eta \rightarrow y} \frac{E_{T}}{N_{\textrm{ch}}} 
\frac{dN_{\textrm{ch}}}{d\eta} \, \, , 
\end{equation}
where $E_T/N_{\textrm{ch}} \equiv \langle dE_T/d\eta \rangle / 
\langle dN_{\textrm{ch}}/d\eta \rangle$ is the average transverse energy density 
per charged track and $C_{\eta \rightarrow y} = 1.25 \pm 0.05$ is a scale factor 
converting $dE_T/d\eta\vert_{\eta = 0}$ to $dE_T/dy\vert_{y = 0}$. 
In Au+Au collisions at \sNN = 200 GeV, $E_T/N_{\textrm{ch}}$ is constant over a 
broad range of event centralities (0-60\%).  We assume that 
$(E_T/N_{\textrm{ch}})_{\textrm{AuAu}}$ is a good approximation of 
$(E_T/N_{\textrm{ch}})_{\textrm{UU}}$.

The most important 
ingredient is the overlap area $S_\perp$, calculated in the Glauber framework.
It depends strongly on the definition used in calculations. For example, the 
values of $S_\perp$ published by PHENIX 
\cite{Adler:2004zn} and STAR \cite{Abelev:2008ez} are a factor of five larger 
than results calculated for 
U+U and Au+Au collisions in Refs.~\cite{PhysRevC.73.034911,Masui2009440}. 
When the relative differences between the energy densities of U+U and Au+Au 
collisions are considered, this discrepancy is unimportant as long as a
consistent definition is used. However it is important when the absolute value 
of $\epsilon_{\textrm{B}}$ is calculated. In this paper, we consider the case 
where $S_\perp$ is defined as the transverse area of the overlap zone weighted 
by the number of participants \cite{PhysRevC.73.034911,Masui2009440}. 
This definition gives the effective ``hot" transverse area of the overlap zone
but neglects areas with low participant density, a rather small effect. 

\begin{figure}[htdp]
\begin{tabular}{ccc}
(a) U+U Tip+Tip &
(b) U+U Side+Side &
(c) U+U Averaged over configurations \\
\includegraphics[trim=0 0 0 1.8cm, clip=true, width=0.33\textwidth]{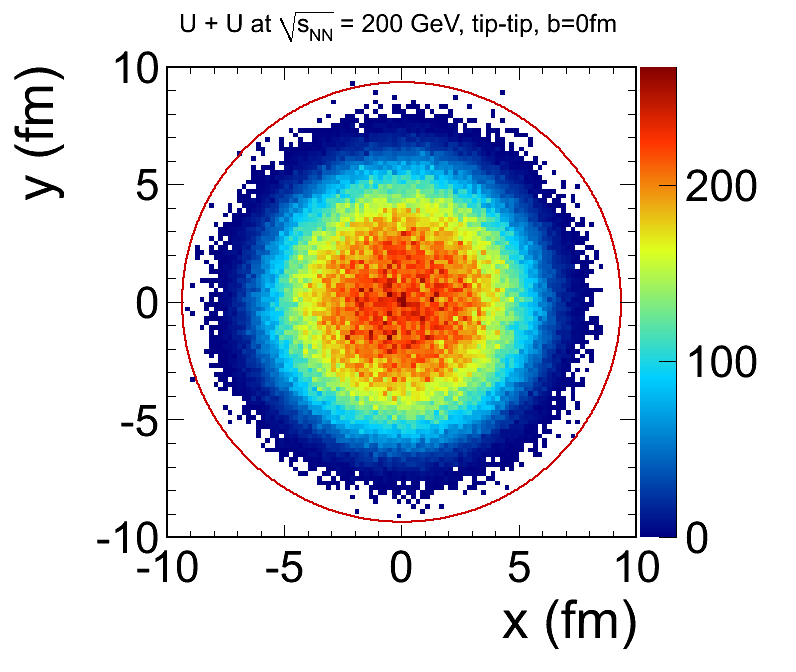} &
\includegraphics[trim=0 0 0 1.8cm, clip=true, width=0.33\textwidth]{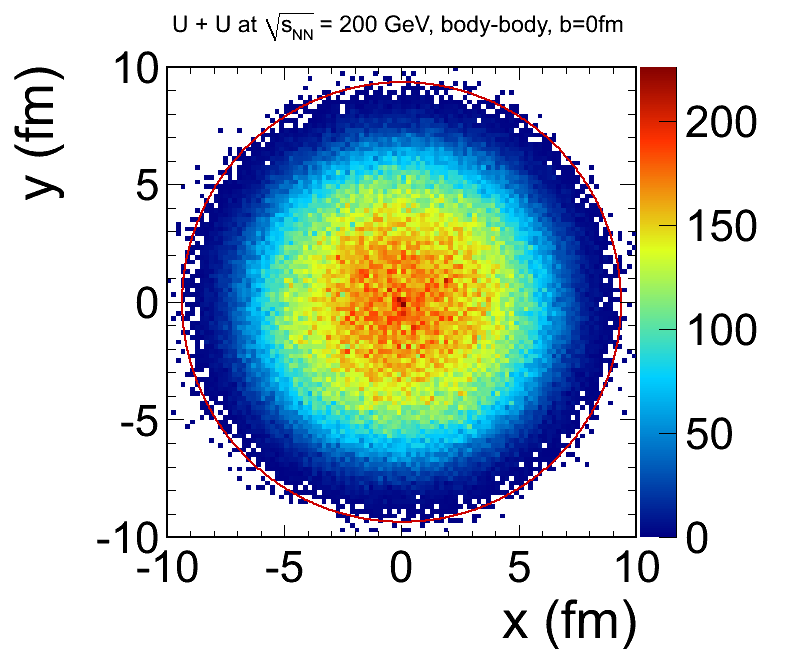} &
\includegraphics[trim=0 0 0 1.8cm, clip=true, width=0.33\textwidth]{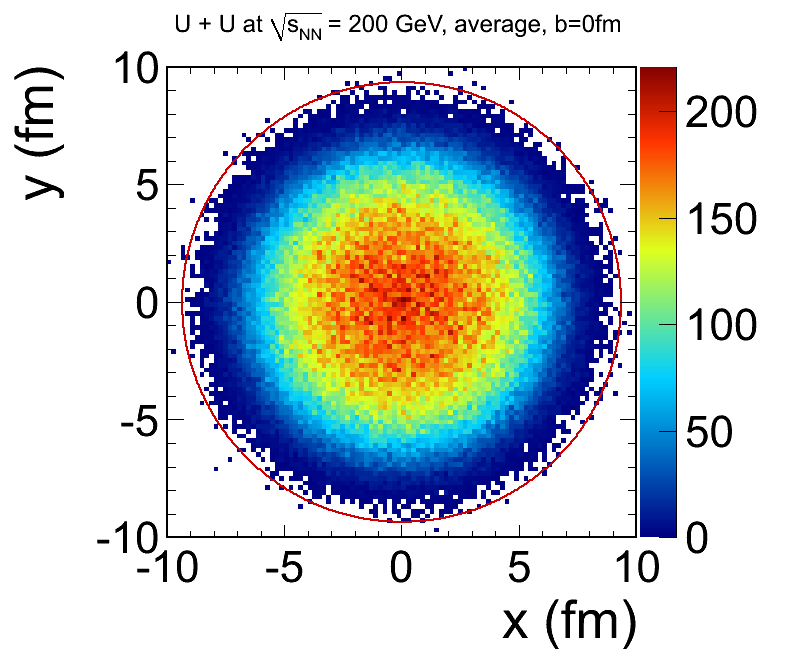} \\ 
\end{tabular}
\caption{\label{Fig:pertDenistyProfileUUb0}(Color online) Participant density 
profiles for U+U collisions in the TT and SS configurations as well as
orientation-averaged collisions at $b = 0$. The circles are shown to guide
the eye and are common to all plots~\cite{Masui2009440,H:Masui:private}.}
\end{figure}

Figure~\ref{Fig:pertDenistyProfileUUb0} shows an example of the participant 
density profile for U+U collisions at $b=0$ in the TT and SS configurations 
as well as averaged over configurations calculated in the Glauber framework of 
Ref.~\cite{Masui2009440}. Much higher density is observed in TT compared to SS 
configurations as well as relative to orientation-averaged U+U collisions 
since the average is dominated by SS configurations \cite{H:Masui:private}. 
Note also the narrower transverse profile of the TT configurations in 
Fig.~\ref{Fig:pertDenistyProfileUUb0}.

\begin{figure}[htdp]
\centering
\includegraphics[width=0.5\textwidth]{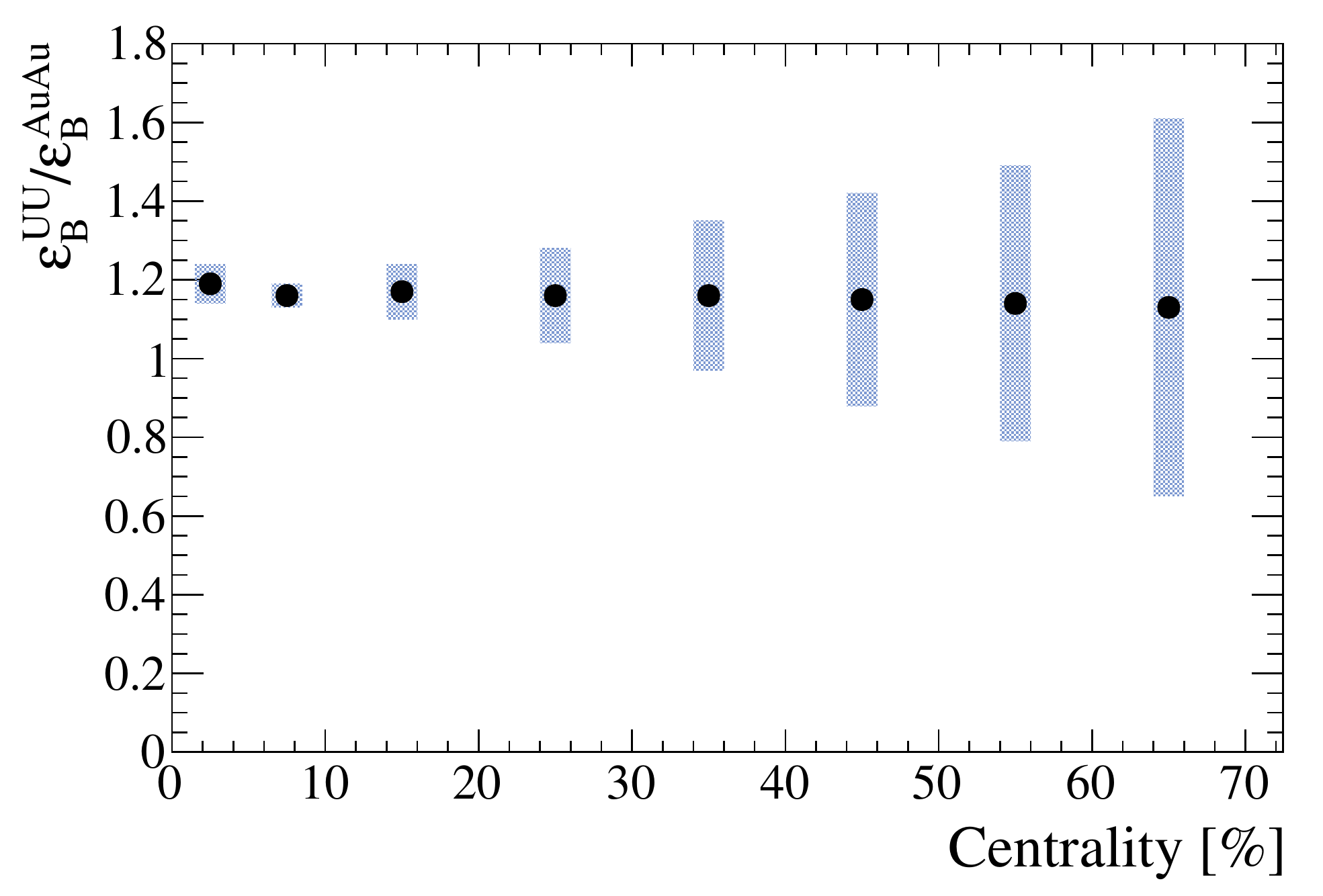} 
\caption{\label{Fig:eBjorkenRatioForAuAuAndUU} The ratio of Bjorken energy 
densities in U+U and Au+Au collisions as a function of centrality. 
The shaded boxes are the systematic uncertainties on $dN^{\rm UU}_{\rm ch}/dy$ and 
 $dN^{\rm AuAu}_{\rm ch}/dy$ added in quadrature. The statistical uncertainties 
are negligible \protect\cite{Masui2009440}.}
\end{figure}

\begin{figure}[htdp]
\begin{tabular}{cc}
\includegraphics[width=0.48\textwidth]{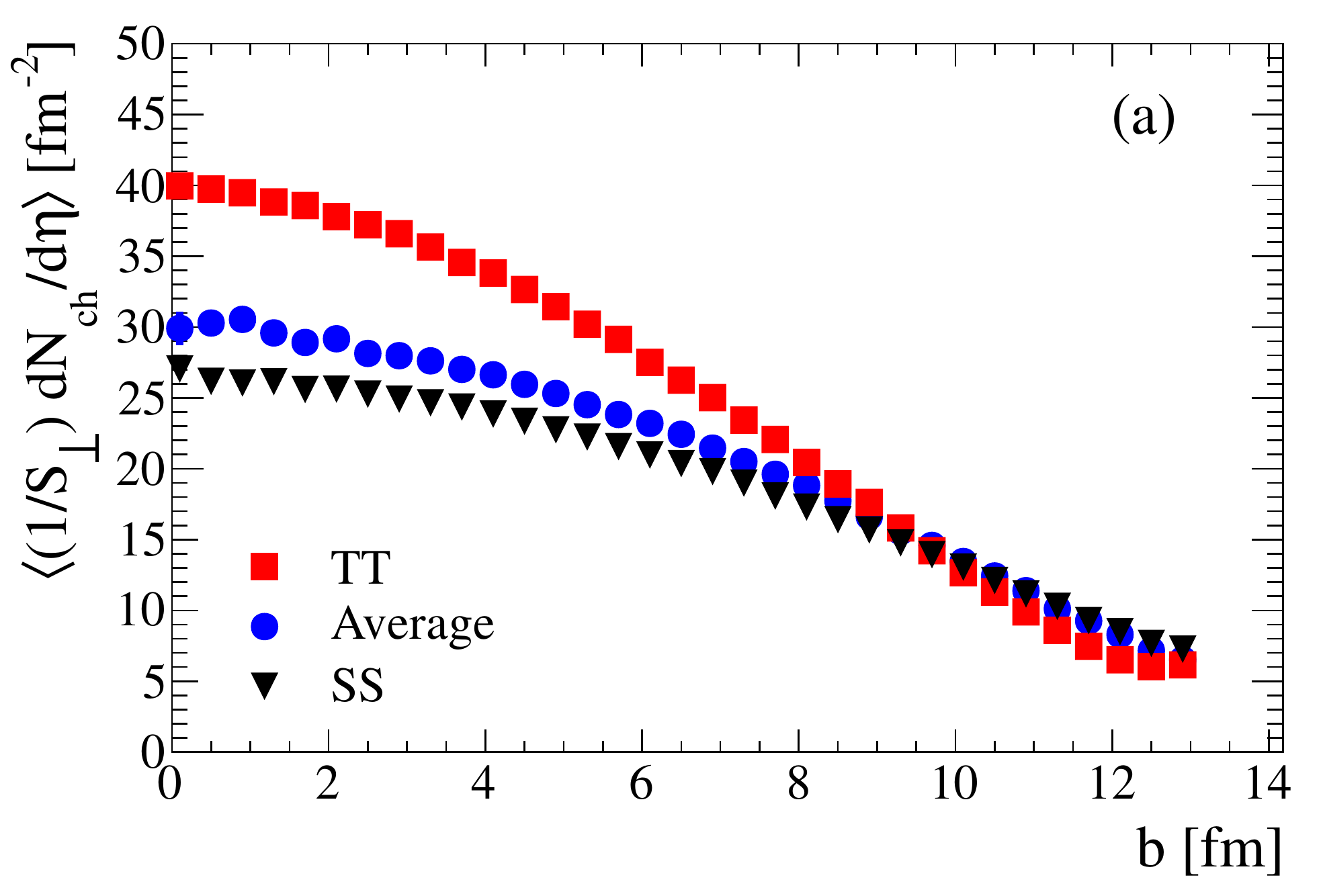} &
\includegraphics[width=0.48\textwidth]{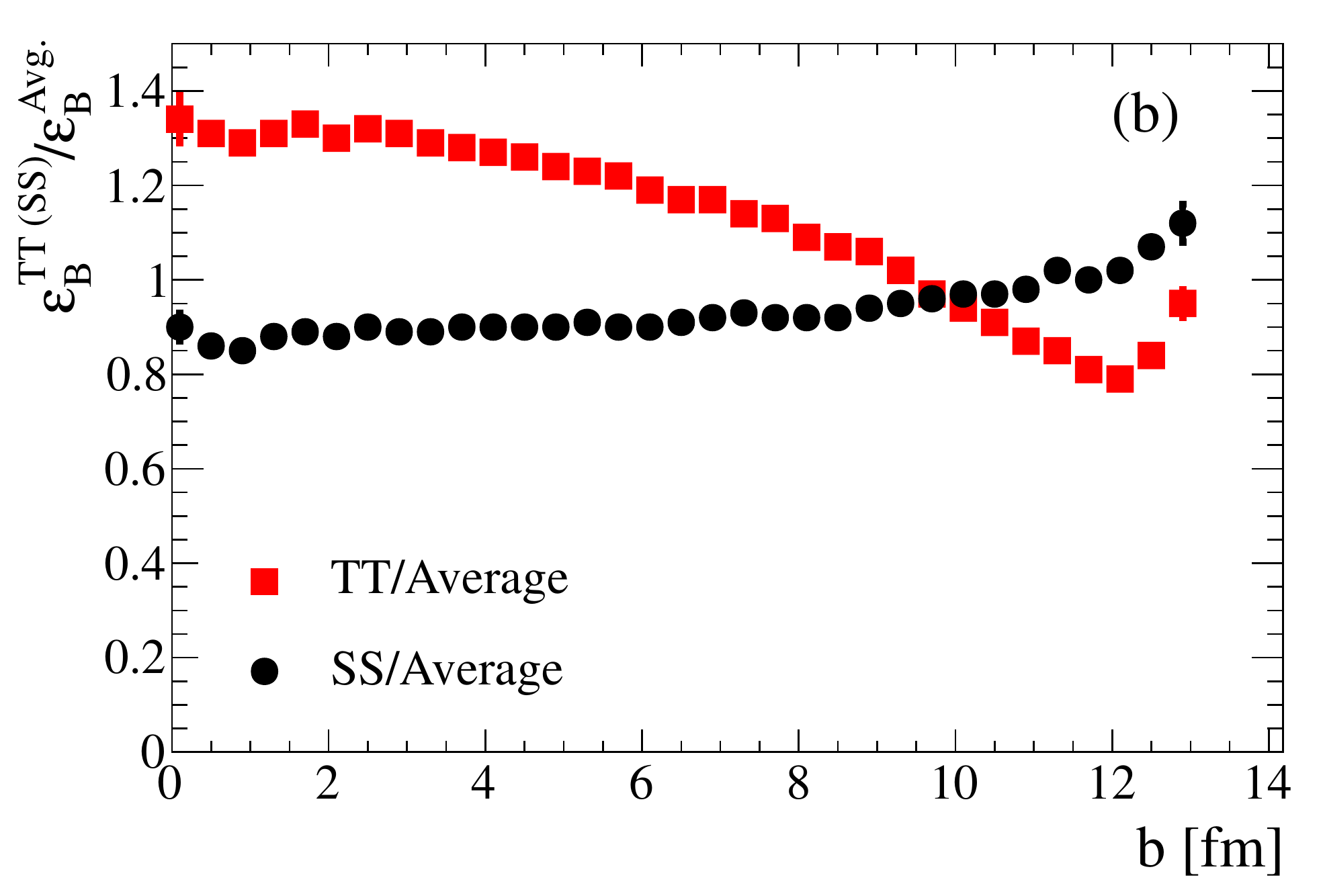} \\ 
\end{tabular}
\caption{\label{Fig:dNdch_UU}(Color online) (a): The average charged track 
density per unit of pseudorapidity $\eta$ in U+U collisions in TT and SS 
configurations as well as orientation-averaged
collisions as a function of impact parameter.  (b):  The ratios of
energy densities in TT and SS orientations relative to orientation-averaged 
collisions as a function of $b$.  In both cases, the statistical uncertainties 
are less than the size of the symbols. 
}
\end{figure}

We use the results of Ref.~\cite{Masui2009440} to estimate the increase in 
energy density in U+U relative to Au+Au collisions. We first estimate 
$\epsilon_{\textrm{B}}$ in orientation-averaged U+U collisions. 
Figure~\ref{Fig:eBjorkenRatioForAuAuAndUU} shows the ratio of the product 
$\epsilon_{\textrm{B}}
\tau$ for orientation-averaged Au+Au and U+U collisions.  We note that
$\epsilon_B$ is $15-20$\% larger in U+U collisions relative to Au+Au 
collisions.  Moreover, U+U collisions in the TT configuration could provide an
increase of up to 30\% in the charged track density, 
$1/S_\perp dN_{\textrm{ch}}/d\eta$, relative to orientation-averaged collisions at 
the same value of $b$, as shown in Fig. \ref{Fig:dNdch_UU}(a). Since 
$1/S_\perp dN_{\textrm{ch}}/d\eta$ is proportional to $\epsilon_B$, TT 
configurations can thus increase $\epsilon_B \tau$ by 20-30\% in central and 
semi-central ($b < 8$ fm) U+U collisions, as shown in 
Fig.~\ref{Fig:dNdch_UU}(b).

In our case, the relevant formation time, $\tau$, is the time required to 
develop the qaurkonia wavefunction.  The lower limit on the charmonium
formation time is the time required for color neutralization of the 
$c\overline{c}$ pair, $\tau_{\psi} \simeq (2m_c \Lambda_{\textrm{QCD}})^{1/2} \simeq 
0.25$~fm for low \pt\ \jpsi\ \cite{Kharzeev1996316}. The formation time of the 
final-state \jpsi, however, may be somewhat longer. 
In Ref.~\cite{Rapp:2008tf}, the formation time for low \pt\ ground state
quarkonia is estimated to be the inverse of the binding energy, $E_{\rm bind}$,
$\tau \sim 1/E_{\textrm{bind}}$.  With 
$E_{\textrm{bind}}^{J/\psi} = 0.64$~GeV and  $E_{\textrm{bind}}^{\Upsilon} = 1.1$~GeV 
\cite{Satz:QuarkoniaBinding}, the formation times for the \jpsi\ and 
$\Upsilon(1S)$ are 
$\tau_{\mathrm{J/\psi}} \approx 0.3$ fm and $\tau_{\mathrm{\Upsilon}} \approx 0.18$ fm 
respectively. We used two values of $\tau$ in our studies: $\tau = 0.25$~fm to 
evaluate the maximum temperature relevant to \jpsi\ suppression due to color 
screening (the lower limit on the \jpsi\ formation time) and $\tau = 1.0$~fm, 
a rather conservative estimate of the time needed for the system to reach 
local thermal equilibrium. 
  
We estimate the maximum energy density in $b=0$ U+U collisions with TT 
orientations. In this case, $1/S_\perp 
dN_{\textrm{ch}}/dy\vert_{y=0} \approx 43.9$ fm$^{-2}$ while $1/S_\perp 
dN_{\textrm{ch}}/dy\vert_{y=0} \approx 31.5$ fm$^{-2}$ in orientation-averaged
Au+Au collisions \cite{UU:experimentalSelection}. A 
proposed experimental method for selecting TT orientations gives $1/S_\perp 
dN_{\textrm{ch}}/dy\vert_{y=0} \approx 40$ fm$^{-2}$ although an effective trigger 
has yet to be developed~\cite{UU:experimentalSelection}. 

In order to convert energy density to temperature and estimate $T/T_c$, we use 
lattice QCD results for $\epsilon(T)$ from Ref.~\cite{PhysRevD.80.014504} with 
$T_c \approx 185$~MeV.  We find, that, for $\tau = 0.25$~fm, the temperature 
exceeds $2.7 T_c$ in both U+U and Au+Au central collisions. However, for 
$\tau = 1.0$~fm, the temperature is below $2T_c$ in Au+Au collisions while it 
could reach $2T_c$ in central U+U collisions averaged over orientations and 
may exceed $2T_c$ in TT orientations, see Table~\ref{Tab:TemperatureUU}. 
Consequently, selecting TT configurations in U+U collisions would extend the 
range of temperatures accessible at RHIC to test the hypothesis of sequential 
quarkonium melting. The temperature in U+U TT collisions would thus be high 
enough to determine whether the $\Upsilon(1S)$ dissociates at $T \approx 2T_c$ 
\cite{Petreczky:Quarkonium}. 
  
\begin{table}[ht]
\caption{\label{Tab:TemperatureUU} The accessible temperature for 
$\tau = 1$~fm in the 5\% most central Au+Au and U+U collisions averaged over 
spatial configurations, ideal TT U+U collisions and those which may be 
selected experimentally, TT (exp), as described in 
Ref.~\protect\cite{UU:experimentalSelection}. }
\begin{ruledtabular}
\begin{tabular}{lcc}
System & $T$ [MeV]  &  $T/T_c$ \\ \hline 
Au+Au (0-5\%) &  351$_{-10} ^{+9}$  &  1.90 \\ 
U+U (0-5\%)  & 366$_{-10} ^{+9}$ & 1.98 \\ 
U+U (TT, exp) & 376 & 2.03 \\ 
U+U (TT, ideal) & 385 & 2.08 \\ 
\end{tabular} 
\end{ruledtabular}
\end{table}

\section{Experimental selection of TT and SS configurations}

There are several strategies which can be used to select TT and SS (or TT- 
and SS-enriched) event samples. In general, the number of binary collisions in 
central and mid-central collisions in TT configurations is higher than in SS
configurations at the same $b$, leading to higher multiplicities in TT
configurations. Therefore TT-enriched head-on ($b=0$) events could be selected
by experimental cuts on the measured charged-track multiplicities. This method 
could be extended to other centralities.  The centrality class can be 
established by employing Zero Degree Calorimeters (ZDCs) which measure the 
energy of spectator nucleons traveling in the forward and backward directions. 
TT-enriched events would be then selected from the centrality class by a 
multiplicity cut. Results obtained in Monte Carlo Glauber 
simulations show that the charged track multiplicity $dN_{\rm{ch}}/d \eta$ in 
central TT collisions is $\approx 17\%$ higher than central collisions in the
SS configuration at $b = 0$.  The difference decreases with $b$ although 
$dN^{\rm TT}_{\rm{ch}}/d \eta$ is still 10\% higher at $b = 4.5$~fm 
\cite{H:Masui:private}. Therefore, the multiplicity difference is suitable 
for offline event selection.  Selection of Tip and Side orientations in $p$U 
collisions would be probably more challenging because the overall multiplicity 
will be lower. However, the number of binary collisions in $p+$Tip interactions 
should be significantly higher than in $p+$Side collisions.  The relative 
difference could be large enough for separation to be feasible in those cases. 

Another approach for selecting collision orientation was proposed in 
Ref.~\cite{UU:experimentalSelection}. TT and SS configurations have different 
elliptical eccentricities and, consequently, different magnitudes of anisotropic
flow. The `reduced flow vector' was proposed in addition to multiplicity cuts 
to select SS and TT samples with high purity. 

In any case, selecting a particular collision geometry requires detailed 
modeling and simulations of experimental observables. 

Note that, for CNM studies, high-purity samples are not 
required: the min-bias data are dominated by SS configurations.  Thus 
TT-enriched samples with sufficiently longer path lengths $L$ than min-bias 
collisions in the same centrality class would be appropriate for studying 
nuclear absorption.  Experimental cuts could thus be adjusted to optimize the 
balance between the desirable $\langle L \rangle$ for physics
and the efficiency of event selection. 

\section{\RUAu: a new observable for \jpsi\ studies}

As shown in Sec.~\ref{Sec:CNM}, the strength of gluon shadowing is
one of the largest uncertainties in the determination of cold nuclear matter
effects on \jpsi\ production.  The effects of shadowing are almost identical 
for Au+Au and U+U collisions averaged over impact parameter.  Thus the relative
uncertainty can be reduced by studying \jpsi\ production in U+U and Au+Au 
collisions within the same centrality class.  As outlined in 
Sec.~\ref{Sec:Intro}, the \jpsi\ yield, $N^{J/\psi}_{\textrm{stat}}$, produced by
statistical recombination is proportional to the number of charm quarks in the 
system: $N^{J/\psi}_{\textrm{stat}} \propto N^2_c$, and $N_{c\overline{c}} \propto 
N_{\textrm{bin}}$, the number of binary nucleon-nucleon collisions. We define the 
relative nuclear modification factor \RUAu\ as
\begin{equation}
R^{\textrm{UU}}_{\textrm{AuAu}} = \frac{dN^{\textrm{UU}}_{J/\psi}/dy}
{dN^{\textrm{AuAu}}_{J/\psi}/dy}  
\bigg( \frac{N_{\textrm{bin}}^{\rm AuAu}}{N_{\textrm{bin}}^{\rm UU}} \bigg)^2 \, \, ,
\end{equation}
the ratio of \jpsi\ yields scaled by the number 
of binary $NN$ collisions, $N_{\textrm{bin}}^2$, in Au+Au and U+U collisions 
respectively. If the \jpsi's are produced by coalescence and suppression is
indepedent of nuclear absorption and energy density, then 
$R^{\rm UU}_{\rm Au+Au} \approx 1$. 

The min bias ratio \RUAu\ has an advantage over \Rcp, the ratio of the \jpsi\ 
yield in 
more central bins relative to that in the most peripheral bin, because \RUAu\ 
does not depend on shadowing, see Figs.~\ref{Fig:Jpsi_RdAu} and
\ref{Fig:Ups_RdAu}.  As we have seen, shadowing can vary 
significantly between central and peripheral collisions but has a rather weak
dependence on impact parameter.  Indeed, if no selection is made on  
orientation in U+U collisions, the shadowing dependence will cancel in the
ratio.

With the large data samples taken by STAR and PHENIX in 2009 and 2010 (and 
even larger data sets expected for RHIC II), the precision of the \jpsi\ yields
will be driven by systematic rather than statistical uncertainties. The 
dominant systematic error ($\pm 10\%$) in \pp\ interactions is the estimate 
of the integrated luminosity \cite{Adare:2006kf} which does not cancel between 
d+Au and Au+Au measurements, as seen in Ref.~\cite{Adare:2010fn} where the 
systematic error on $R_{\textrm{dAu}}$ is much larger than that on \Rcp. In the 
case of \RUAu, systematic uncertainties would mostly cancel when Au+Au and U+U 
data are taken by the same detector.

These features of \RUAu\ make it useful for testing models of \jpsi\ production 
and in-medium interactions as described in Sec.~\ref{Sec:Intro}. The dependence
of nuclear absorption on \bb\ is rather small, but can be 
detected if large data sets are collected. Furthermore \RUAu$(y)$ will provide 
an additional test of models which reproduce $R_{\textrm{AuAu}}(y)$.

\section{Summary}

We have investigated cold nuclear matter effects on charmonium production in 
U+U collisions. Such collisions provide an interesting opportunity to study 
\jpsi\  in-medium interactions since model-dependent uncertainties can be 
significantly reduced. We propose a new observable, \RUAu, which is free from 
some of the uncertainties associated with $R_{{\rm AuAu}}$ and \Rcp. The 
experimental techniques for quarkonium measurements at RHIC are well 
established; there are abundant Au+Au data; and U+U collisions are planned 
at RHIC in the near future. Therefore, our proposed observable \RUAu\ represents
an additional, important handle on \jpsi\ in-medium interactions. 
Moreover, the energy 
density achievable in U+U relative to Au+Au collisions is up to $\approx 20\%$ 
larger, extending the range of energy densities available for testing 
quarkonium suppression due to color screening. 

Furthermore, U+U collisions would be also useful for cold nuclear matter 
studies at lower energies, as part of a RHIC Beam Energy Scan program. Nuclear 
absorption is expected to be larger at lower energies \cite{CarlosHermineme} 
while there may be antishadowing of the nuclear gluon distribution rather
than shadowing. Therefore cold nuclear matter effects could be tested
more directly by lower energy U+U collisions.

\acknowledgments{}

We thank Hiroshi Masui for providing the Glauber calculations of 
Ref.~\cite{Masui2009440}. This work was performed under the auspices of the 
U.S. Department of Energy by Lawrence Livermore National Laboratory under 
Contract DE-AC52-07NA27344 (RV), by Lawrence
Berkeley National Laboratory under Contract DE-AC02-05CH11231 (GO and DK) 
and was also supported in part by the National 
Science Foundation Grant NSF PHY-0555660 (RV).



\begin{thebibliography}{99} 
\bibliographystyle{apsrev}

\bibitem{Heinz:2004ir}
U.~W.~Heinz and A.~Kuhlman,
Phys.\ Rev.\ Lett.\  {\bf 94}, 132301 (2005).

\bibitem{PhysRevC.73.034911}
C.~Nepali, G.~Fai and D.~Keane, Phys.\ Rev.\ C {\bf 73}, 034911 (2006)
  
\bibitem{Masui2009440}
H.~Masui, B.~Mohanty and N.~Xu,
Phys.\ Lett.\  B {\bf 679}, 440 (2009).
  
\bibitem{Hirano:2010jg}
T.~Hirano, P.~Huovinen and Y.~Nara,
Phys.\ Rev.\  C {\bf 83}, 021902 (2011).

\bibitem{Voloshin:1999gs}
S.~A.~Voloshin and A.~M.~Poskanzer,
Phys.\ Lett.\  B {\bf 474}, 27 (2000).

\bibitem{STAR:Flow}
B.~I.~Abelev {\it et al.}  (STAR Collaboration),
Phys.\ Rev.\  C {\bf 77}, 054901 (2008).
  
\bibitem{Kolb:2000sd}
P.~F.~Kolb, J.~Sollfrank and U.~W.~Heinz,
Phys.\ Rev.\  C {\bf 62}, 054909 (2000).  
  
\bibitem{Matsui1986416}
T.~Matsui and H.~Satz, 
Phys.~Lett.~B  {\bf 178} (1986) 416.
	
\bibitem{NA501997}
M.~C.~Abreu it et al. (NA50 Collaboration),
Phys.\ Lett.\  B {\bf 410}, 327 (1997). 
  
\bibitem{NA50PbPb}
B. Alessandro {\it et al.} (NA50 Collaboration), Eur. Phys.
J. C {\bf 39}, 335 (2005).
  
\bibitem{Na602007}
R.~Arnaldi et al. (NA60 Collaboration),
Phys.\ Rev.\ Lett. {\bf 99}, 132302 (2007).	

\bibitem{PhenixJpsiAuAu}
A.~Adare {\it et al.}  (PHENIX Collaboration),
Phys.\ Rev.\ Lett.\  {\bf 98}, 232301 (2007).

\bibitem{Kluberg:2009wc}
L.~Kluberg and H.~Satz,
arXiv:0901.3831 [hep-ph].
  
\bibitem{Rapp:2008tf}
R.~Rapp, D.~Blaschke and P.~Crochet,
Prog.\ Part.\ Nucl.\ Phys.\  {\bf 65}, 209 (2010).
  
\bibitem{RVgeneral}
R.~Vogt, Phys.~Rev.~C {\bf 71}, 054902 (2005).
  
\bibitem{EPS09}
K. J. Eskola, H. Paukkunen and C. A. Salgado, JHEP
{\bf 0904}, 065 (2009).
	
\bibitem{NA50}
B. Alessandro {\it et al.} (NA50 Collaboration), Eur. Phys.
J. C {\bf 33}, 31 (2004); ibid. {\bf 48}, 329 (2006).

\bibitem{E866}
M. J. Leitch {\it et al.} (E866 Collaboration), Phys. Rev. Lett.
{\bf 84}, 3256 (2000).	  
  
\bibitem{RVabschi}
R.~Vogt, Nucl.~Phys.~A {\bf 700}, 539 (2002).

\bibitem{Karsch:2005nk}
F.~Karsch, D.~Kharzeev and H.~Satz,
Phys.\ Lett.\  B {\bf 637}, 75 (2006).

\bibitem{RVSG}
S. Gavin and R. Vogt, Nucl. Phys. B {\bf 345}, 104 (1990). 

\bibitem{RVPhysRept}
R. Vogt, Phys. Rept. {\bf 310}, 197 (1999).

\bibitem{Capella}
A.~Capella, L. Bravina, E.~G. Ferreiro, A.~B. Kaidalov, K. Tywoniuk and
E. Zabrodin, Eur. Phys. J. C {\bf 58}, 437 (2008).

\bibitem{HSD}
W.~Cassing, E.~L.~Bratkovskaya, and S.~Juchem, Nucl.~Phys.~A {\bf 674}, 249 
(2000).

\bibitem{ThewsMangano}
R.~L.~Thews and M.~L.~Mangano, Phys.~Rev.~C {\bf 73}, 014904 (2006).

\bibitem{statistical}
A.~Andronic, P.~Braun-Munzinger, K.~Redlich and J.~Stachel, Phys.~Lett.~B 
{\bf 571}, 36 (2003);
A.~P.~Kostyuk, M.~I.~Gorenstein, H.~St\"ocker and W.~Greiner, Phys.~Rev.~C 
{\bf 68}, 041902 (2003).

\bibitem{RVTFTU}
A.~D.~Frawley, T.~Ullrich and R.~Vogt,
Phys.\ Rept.\  {\bf 462}, 125 (2008).

\bibitem{Vogt:2010aa}
R.~Vogt,
Phys.\ Rev.\  C {\bf 81}, 044903 (2010).

\bibitem{SpatialShadDep1999}
V.~Emel'yanov, A.~Khodinov, S.~R.~Klein and R.~Vogt,
Nucl.\ Phys.\  A {\bf 661}, 649 (1999).

\bibitem{Inhomogeneous:Shadowing}
S.~R.~Klein and R.~Vogt,
Phys.\ Rev.\ Lett.\  {\bf 91}, 142301 (2003).

\bibitem{rvhip}
R. Vogt, Heavy Ion Phys. {\bf 25} (2006), 97.

\bibitem{tuchin}
D. Kharzeev, E. Levin, M. Nardi and K. Tuchin, Nucl. Phys. 
A {\bf 826}, 230 (2009).

\bibitem{boris}
B.~Z.~Kopeliovich {\it et al}, Phys. Rev. C {\bf 83}, 014912 (2011).
  
\bibitem{CarlosHermineme}
C.~Lourenco, R.~Vogt and H.~K.~Woehri,
JHEP {\bf 0902} (2009) 014.    

\bibitem{QWGdoc}
N. Brambilla {\it et al}, Eur. Phys. J. C {\bf 71} (2011) 1.
  
\bibitem{Adare:2010fn}
A.~Adare {\it et al.}, (PHENIX Collaboration)
arXiv:1010.1246 [nucl-ex].

\bibitem{Jamie}
J. L.~Nagle, A. D. Frawley, L. A. Linden Levy and M. G. Wysocki, 
arXiv:1011.4534 [nucl-th]

\bibitem{RVDMTF}
A.~D.~Frawley, D.~McGlinchey and R.~Vogt, in preparation.

\bibitem{Adare:2008sh}
  A.~Adare {\it et al.}  [PHENIX Collaboration],
  Phys.\ Rev.\ Lett.\  {\bf 101}, 122301 (2008)


\bibitem{Upsilon:dAu}
H.~Liu  (STAR Collaboration),
Nucl.\ Phys.\  A {\bf 830}, 235c (2009).  

\bibitem{lansberg}    
E. G. Ferreiro, F. Fleuret, J.-P. Lansberg and A.
Rakotozafindrabe, Phys. Lett. B {\bf 680}, 50 (2009).

\bibitem{STAR:Upsilon:HP2010}
R.~Reed (STAR Collaboration), 
in proceedings of the 4$^{\rm th}$ International Conference on Hard and 
Electromagnetic Probes of High-Energy Nuclear Collisions (HP2010) (2010).
    
\bibitem{Bjorken:1982qr}
J.~D.~Bjorken,
Phys.\ Rev.\  D {\bf 27}, 140 (1983).    

\bibitem{Adler:2004zn}
S.~S.~Adler {\it et al.}  (PHENIX Collaboration),
Phys.\ Rev.\  C {\bf 71}, 034908 (2005) [Erratum-ibid.\  C {\bf 71}, 049901 (2005)].
 
\bibitem{Abelev:2008ez}
B.~I.~Abelev {\it et al.}  (STAR Collaboration),
Phys.\ Rev.\  C {\bf 79}, 034909 (2009). 

\bibitem{H:Masui:private}
H.~Masui, private~communications.
 
\bibitem{Kharzeev1996316}
D.~Kharzeev and H.~Satz,
Phys.\ Lett.\  B {\bf 366}, 316 (1996).

\bibitem{Satz:QuarkoniaBinding}
H.~Satz,
J.\ Phys.\ G {\bf 32}, R25 (2006).  
 
\bibitem{UU:experimentalSelection}
C.~Nepali, G.~I.~Fai and D.~Keane,
Phys.\ Rev.\  C {\bf 76}, 051902 (2007)
[Erratum-ibid.\  C {\bf 76}, 069903 (2007)]. 
      
\bibitem{PhysRevD.80.014504}
A.~Bazavov {\it et al.},
Phys.\ Rev.\  D {\bf 80}, 014504 (2009).

\bibitem{Petreczky:Quarkonium}
A.~Mocsy and P.~Petreczky,
Phys.\ Rev.\ Lett.\  {\bf 99}, 211602 (2007).

\bibitem{Adare:2006kf}
A.~Adare {\it et al.}  [PHENIX Collaboration], 
Phys.\ Rev.\ Lett.\  {\bf 98}, 232002 (2007).  
    
\end{thebibliography}
\end{document}